\documentclass[conference]{IEEEtran}
\IEEEoverridecommandlockouts
\usepackage{cite}
\usepackage{amsmath,amssymb,amsfonts}
\usepackage{algorithmic}
\usepackage{textcomp}
\usepackage{xcolor}
\usepackage{microtype}
\usepackage{subcaption}
\usepackage{booktabs} 
\usepackage{graphicx}
\usepackage{multirow}
\usepackage{url}
\usepackage{hyperref}
\usepackage{float}

\def\BibTeX{{\rm B\kern-.05em{\sc i\kern-.025em b}\kern-.08em
    T\kern-.1667em\lower.7ex\hbox{E}\kern-.125emX}}
\begin{document}

\title{Dissecting the Runtime Performance of the Training, Fine-tuning, and Inference of Large Language Models\\

\thanks{*Equal contribution.}
\thanks{{\dag}Corresponding author.}
}

\author{\IEEEauthorblockN{Longteng Zhang$^{1*}$, Xiang Liu$^{1*}$, Zeyu Li$^{1}$, Xinglin Pan$^{1}$, Peijie Dong$^{1}$, Ruibo Fan$^{1}$}
\IEEEauthorblockN{Rui Guo$^{2}$, Xin Wang$^{2}$, Qiong Luo$^{1,4}$, Shaohuai Shi$^{3\dag}$, Xiaowen Chu$^{1,4\dag}$}
\IEEEauthorblockA{
\textit{$^{1}$The Hong Kong University of Science and Technology (Guangzhou)} \\
\textit{\{lzhang330, xliu886, zli755, xpan413, pdong212, rfan404\}@connect.hkust-gz.edu.cn} \\
\textit{$^{2}$Beijing Damo Technology Co. Ltd. greatfeel@gmail.com, wangxin@aichoice.cn} \\
\textit{$^{3}$Harbin Institute of Technology, Shenzhen shaohuais@hit.edu.cn} \\
\textit{$^{4}$The Hong Kong University of Science and Technology luo@cse.ust.hk, xwchu@ust.hk}} \\
}

\maketitle

\begin{abstract}
Large Language Models (LLMs) have seen great advance in both academia and industry, and their popularity results in numerous open-source frameworks and techniques in accelerating LLM pre-training, fine-tuning, and inference. Training and deploying LLMs are expensive as it requires considerable computing resources and memory, hence many efficient approaches have been developed for improving system pipelines as well as operators. However, the runtime performance can vary significantly across hardware and software stacks, which makes it difficult to choose the best configuration. In this work, we aim to benchmark the performance from both macro and micro perspectives. First, we benchmark the end-to-end performance of pre-training, fine-tuning, and serving LLMs in different sizes , i.e., 7, 13, and 70 billion parameters (7B, 13B, and 70B) on three 8-GPU platforms with and without individual optimization techniques, including ZeRO, quantization, recomputation, FlashAttention. Then, we dive deeper to provide a detailed runtime analysis of the sub-modules, including computing and communication operators in LLMs. For end users, our benchmark and findings help better understand different optimization techniques, training and inference frameworks, together with hardware platforms in choosing configurations for deploying LLMs. For researchers, our in-depth module-wise analyses discover potential opportunities for future work to further optimize the runtime performance of LLMs.
\end{abstract}


%

\section{Introduction}\label{sec:introduction}
In recent years, large language models (LLMs) have become very popular in AI applications~\cite{ouyang2022training,chang2023survey}. With an increased size, LLMs demonstrate much better generalization capabilities in various tasks~\cite{kaplan2020scaling,hoffmann2022training,openai2023gpt4}. However, the model size gets huge in recent work, for example, GPT-3~\cite{brown2020language} has 175 billion parameters and PaLM~\cite{chowdhery2022palm} has 540 billion parameters. As a result, training and deploying LLMs is complex and expensive.

\begin{figure}[!t]
	\centering
\includegraphics[width=\linewidth]{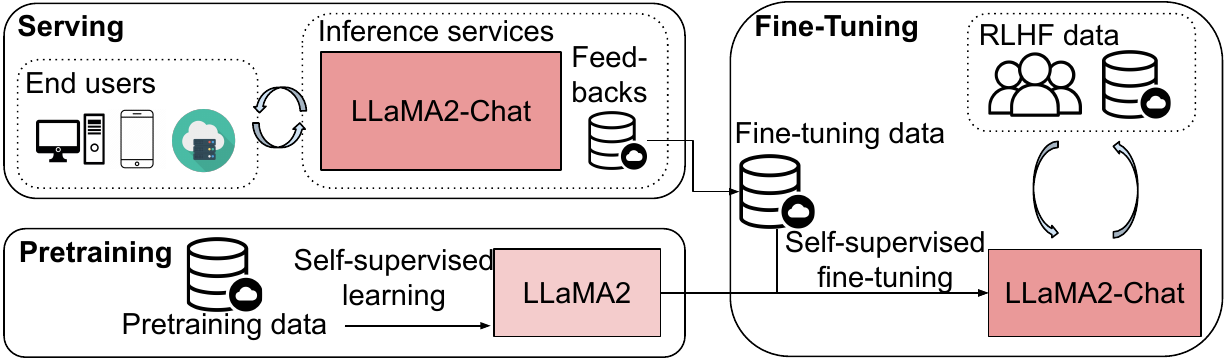}
	\caption{The pipeline for deploying an LLM using Llama2~\cite{touvron2023llama_2} and Llama2-Chat as an example, which contains three main stages: pre-training, fine-tuning, and serving.}
	\label{fig:pipeline}
\vspace{-20px}
\end{figure}

Specifically, the pipeline of LLMs (as shown in Figure~\ref{fig:pipeline}) has three main stages, pre-training, fine-tuning, and serving, for deploying a LLM for a real-world application~\cite{kaddour2023challenges}. First, the model (e.g., Llama2) is pre-trained using self-supervised learning before it is applied to downstream tasks, which is the most time-consuming stage in the LLM pipeline. For example, pre-training a PaLM model requires around $2.5\times 10^{24}$ floating-point operations (FLOPs) and takes 64 days when executed on 6,144 Google TPUv4 chips~\cite{chowdhery2022palm}. Second, the pre-trained model is further fine-tuned on downstream tasks or instruction datasets\footnote{Instruction tuning with human feedback is also regarded as fine-tuning throughout this paper as its paradigm is almost identical with fine-tuning with downstream task datasets.} to improve its performance in real-world applications~\cite{ouyang2022training}, e.g., Llama2-Chat is fine-tuned with Llama2 using fine-turning and RLHF data. Third, after the model has been fine-tuned (e.g., Llama2-Chat), it is deployed as a web (or API) service that provides inference results for given input queries.

To reduce the computational costs in the pipeline of LLMs, dedicated frameworks have been proposed for efficient pre-training (e.g., DeepSpeed~\cite{rasley2020deepspeed} and Megatron-LM~\cite{narayanan2021efficient}), fine-tuning (e.g., PEFT~\cite{peft}), and inference (e.g., vLLM~\cite{kwon2023efficient}, LightLLM~\cite{LightLLM2023}, and TGI~\cite{TGI2023}). Inside each framework, optimization techniques have been applied for memory and compute efficiency. Specifically, in pre-training, memory-efficient approaches (ZeRO~\cite{rajbhandari2020zero}, activation recomputation~\cite{korthikanti2023reducing,jain2020checkmate,smith2022using}, and quantization~\cite{yao2022zeroquant}) are commonly adopted to enable GPUs that have limited memory to train large models. In fine-tuning, parameter-efficient fine-tuning (PEFT) methods such as LoRA~\cite{hu2022lora,dettmers2023qlora} have been used to fine-tune LLMs by tuning the parameters of adaptors instead of full parameters of the model so that GPUs with limited memory are able to fine-tune LLMs. In LLM serving, to maximally utilize the GPU resources in deployment, quantization~\cite{yao2022zeroquant} and kernel-level optimization~\cite{dao2022flashattention} are performed on the trained model.

However, with various LLM frameworks and related optimization techniques running on different types of hardware (e.g., high-end GPUs such as Nvidia A800 and consumer-level GPUs, e.g., Nvidia Geforce RTX4090 and RTX3090), there are two under-explored problems that are crucial for end-users and researchers. First, \textit{what configurations are required for pre-training, fine-tuning, and deploying LLMs for particular applications to balance the efficiency and cost?} For example, are 8x A800 80GB GPUs sufficient to pre-train a 7B model, how long will it take, and what kind of optimization techniques should be enabled to accelerate the training? Second, \textit{do existing state-of-the-art systems with highly optimized techniques fully utilize the GPU resources and where is the performance bottleneck?} In particular, what is the peak utilization of compute and bandwidth resources on modern GPU servers under different configurations? 

To address these questions, we benchmark the runtime and memory performance of existing systems in the LLM pipeline on various types of GPU servers. Specifically, we provide the following detailed benchmarks to understand the time and memory efficiency of different software and hardware systems. (1) On the framework-level, we choose DeepSpeed and Megatron-LM to study their training performance of Llama2~\cite{touvron2023llama_2} with three scales (7B, 13B, and 70B) on three types of hardware (A800, RTX4090, and RTX3090 servers). (2) We study the impact on the memory and compute efficiency of integrating ZeRO, quantization, activation recomputation, and FlashAttention. (3) We evaluate popular PEFT frameworks including LoRA and QLoRA, to understand their fine-tuning efficiency. (4) We study the end-to-end inference performance using highly optimized inference libraries including vLLM, LightLLM, and TGI. (5) To understand the performance in-depth, we microbenchmark the key kernels that are the most time-consuming.

Through comprehensive benchmarks and analysis, we conclude the following important findings. (1) DeepSpeed achieves higher throughput than Megatron-LM in all configurations. (2) ZeRO saves great amount of memory without sacrificing training efficiency, or it may suffer from OOM when the number of GPUs is under 4. (3) Offloading further reduces memory usage but significantly slows down the training process. (4) Activation recomputation works well only when combined with other optimization techniques, otherwise it cannot reduce much memory consumption. (5) Quantization boosts the training speed, achieves the largest throughput on all hardware platforms, compared with other methods. However it may lead to convergence failure.
(6) FlashAttentionn accelerates the training process on various hardware platforms, with a slightly higher peak memory consumption, which can be migrated with other memory-efficient methods. (7) The PEFT method has enabled various devices to train LLMs. (8) On the A800 platform, LightLLM exhibits superior throughput. In converse, on the 24G GPU platform, TGI demonstrates enhanced throughput, whereas the vLLM and LightLLM display comparable levels of throughput.


\section{Background and Preliminaries}\label{sec:background}
\subsection{Decoder-only Transformer-based LLMs}

\begin{figure}[H]
	\centering
\includegraphics[width=\linewidth]{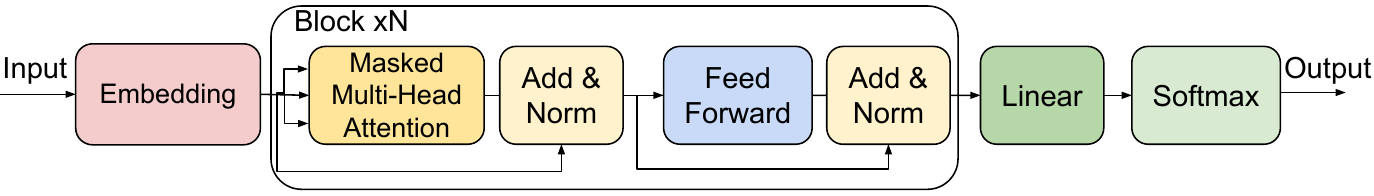}
	\caption{The decoder structure of decoder-only LLMs.}
	\label{fig:decoder}
\end{figure}

The traditional transformer~\cite{vaswani2017attention} consists of encoder and decoder architectures, and the decoder has been widely used for modern text-generation LLMs (e.g., GPT-3~\cite{GPT3}, Llama~\cite{touvron2023llama_1}, Llama2~\cite{touvron2023llama_2}, BLOOM~\cite{scao2022bloom}, etc.). The decoder has the structure as shown in Figure~\ref{fig:decoder}. The input data is first encoded through the embedding layer, whose output are fed into multiple attention blocks. Each attention block consists of a multi-head attention and a feed-forward network which has several linear layers. Then, the output of multiple blocks are concatenated as the input for the next linear layer, called generation or classification head, followed by a softmax layer to calculate the probability for the next token.

\subsection{Pre-training Frameworks}
\textbf{DeepSpeed.} DeepSpeed~\cite{rasley2020deepspeed} is a cutting-edge deep learning (DL) optimization software suite developed for both large-scale training and inference. It adopts ZeRO~\cite{rajbhandari2020zero,ren2021zerooffload,rajbhandari2021zeroinfinity}, offloading, and DeepSpeed-Inference~\cite{aminabadi2022deepspeed}, and other techniques. The software is encapsulated in an open-source library, allowing seamless integration into training and inference. It has been widely adopted in the DL community and is a cornerstone of Microsoft’s AI at Scale initiative.

\textbf{Megatron-LM.} Megatron-LM~\cite{shoeybi2019megatron,narayanan2021efficient} addresses the challenge of efficiently training expensive transformer models. Megatron-LM is well optimized for supporting 3D-parallelism and activation recomputation~\cite{korthikanti2023reducing}. It also introduces sequence parallelism to be combined with tensor parallelism, so it substantially reduces the need for activation recomputation. Since Megatron-LM is highly scalable, it has been a commonly used LLM training system.




\subsection{Fine-tuning Frameworks}
A straight-forward way of fine-tuning LLM for downstream tasks is to fine-tune all parameters (i.e., Full-FT), but it is very memory-expensive and time-consuming. In real-world applications, parameter-efficient fine-tuning (PEFT)~\cite{peft} approaches are more popular as they require much less memory resources to fine-tune a model than Full-FT. In PEFT, LoRA~\cite{hu2022lora} (or QLoRA~\cite{dettmers2023qlora}, the quantization version of LoRA) and Prompt Tuning~\cite{lester2021power} are two widely adopted methods. 

\textbf{LoRA.} The low-rank adaptation (LoRA) method~\cite{hu2022lora} is based on the observation that over-parameterized models often operate within a low intrinsic dimension. For a pre-trained weight matrix \(W_0 \in \mathbb{R}^{d \times k}\), its update is constrained using a low-rank decomposition: \(W_0 + \Delta W = W_0 + BA\), where \(B \in \mathbb{R}^{d \times r}\), \(A \in \mathbb{R}^{r \times k}\), and the rank \(r \ll \min(d, k)\). During training, \(W_0\) remains static without gradient updates, while \(A\) and \(B\) are trainable. The modified forward pass is represented as \(h = W_0x + \Delta Wx = W_0x + BAx\). Thus, LoRA approximates Full-FT by setting the LoRA rank \(r\) to the pre-trained weight matrices to train the low-rank matrices and incurs little additional overhead during inference.




\textbf{QLoRA.} QLoRA~\cite{dettmers2023qlora} is a quantization version of LoRA. It converts pretrained models to a particular 4-bit data type (NormalFloat or NF4), thereby drastically reducing memory usage and improving compute efficiency while preserving data integrity during quantization.


\textbf{Prompt Tuning.} Prompt tuning~\cite{lester2021power} is a novel technique tailored for adapting frozen language models to specific downstream tasks. Specifically, prompt-tuning emphasizes the learning of ``soft prompts'' via backpropagation, allowing them to be fine-tuned using signals from labeled examples. 

\subsection{Inference Frameworks}

\textbf{TGI.} Text Generation Inference (TGI)~\cite{TGI2023} is a toolkit designed specifically for the deployment and serving of LLMs. Catering to a range of renowned open-source LLMs, including Llama series~\cite{touvron2023llama_1,touvron2023llama_2}, BLOOM~\cite{scao2022bloom}. It adopts tensor parallelism (or model parallelism~\cite{dean2012large}) for accelerated inference across multiple GPUs and employs token streaming via Server-Sent Events (SSE)~\cite{vinoski2012server}. Notably, TGI's continuous batching optimizes the handling of incoming requests, maximizing throughput. The toolkit further refines inference with optimized transformer code, leveraging advanced techniques such as FlashAttention~\cite{dao2022flashattention} and PagedAttention~\cite{kwon2023efficient}. 

\textbf{vLLM.} The PagedAttention algorithm~\cite{kwon2023efficient}, inspired by virtual memory and paging mechanisms from operating systems, segments the dynamically changing key-value cache (KV cache) memory into smaller blocks, which can be placed in non-contiguous areas. This approach resolves issues such as fragmentation, paving the way for optimized memory utilization. Building upon the PagedAttention foundation, vLLM\footnote{\url{https://github.com/vllm-project/vllm}} is a high-throughput LLM serving engine. 

\textbf{LightLLM.} LightLLM~\cite{LightLLM2023} stands out as a cutting-edge, Python-based LLM inference and serving framework, distinguished by its lightweight architecture, scalability, and swift performance. LightLLM employs a tri-process asynchronous collaboration, scheme allowing tokenization, model inference, and detokenization occuring concurrently, thereby maximizing GPU utilization. Additionally, it introduces the "Nopad" feature to adeptly manage requests with varying lengths and a dynamic batch scheduling mechanism to streamline request processing. LightLLM's unique token-wise KV cache memory management, termed "Token Attention", significantly reduces memory usage during inference. Another feature is the Int8KV Cache, which effectively doubles the token capacity. 


\subsection{Optimization Techniques}

\textbf{ZeRO.} ZeRO serial techniques (i.e., ZoRO-1/2/3~\cite{rajbhandari2020zero}, ZeRO-Offload~\cite{ren2021zerooffload}, and ZeRO-Infinity~\cite{rajbhandari2021zeroinfinity}) optimize the memory efficiency in training LLMs. ZeRO-1 partitions the model's optimizer states across GPUs, reducing memory used for these states. ZeRO-2 extends ZeRO-1 by adding partitioning of gradients across GPUs, further decreasing memory required for gradients. However, ZeRO-2 introduces extra Reduce collective communication primitives into the backward process. Based on ZeRO-1 and ZeRO-2, ZeRO-3 further adds model parameter partitioning and model parallelism for activations, maximizing memory savings and allowing for the training of even larger models with reduced communication overhead, but it requires extra Reduce-Scatter for partitioning model parameters. With ZeRO-Offload, the authors aim to make billion-scale model training more accessible, bridging the gap between computational demands and available resources.


\textbf{Activation Offloading.} Activation offloading~\cite{ren2021zerooffload} is a technique aimed at efficiently managing the substantial computational and GPU memory demands inherent in training and deploying LLMs. By selectively transferring the activations (intermediate neuron output) from GPU memory to CPU memory or disk storage during the forward pass of a neural network and subsequently reloading them during the backward pass for gradient computations, activation offloading facilitates memory and computational resource optimization. Furthermore, two primary methods, optimizer offloading and model parameter offloading, can be employed to significantly alleviate pressure on GPU memory. However, it introduces additional data transfer overhead.


\textbf{Activation Recomputation.} It involves the re-computation of intermediate activations during the backward pass of training, rather than retaining them from the forward pass so as to optimize memory usage. By eschewing the storage of activations for each layer of a model, it can significantly reduce memory consumption. However, this method introduces additional computational overhead. While the memory benefits are substantial, the recomputation process necessitates alterations to the conventional backpropagation algorithm, adding a layer of complexity to the training paradigm.

\textbf{Quantization.}Quantization is an important technique to represent the weights or activations using low-bit format to reduce both memory size and compute time. ZeroQuant~\cite{yao2022zeroquant} is one of the popular systems, which introduces a novel post-training quantization approach and develops a hardware-friendly quantization scheme for both weights and activations, a unique layer-by-layer knowledge distillation algorithm, and a highly optimized system backend for quantization. It has demonstrated the capability to reduce the precision for weights and activations to INT8 for models such as BERT and GPT-3 with minimal accuracy degradation. 

\textbf{FlashAttention.} FlashAttention~\cite{dao2022flashattention}, is tailored to address the inherent challenges posed by transformers in processing extensive sequences. This algorithm is IO-aware in that it optimizes the interplay between GPU memory levels. It utilizes tiling to reduce memory reads/writes between the GPU's High Bandwidth Memory (HBM) and on-chip Static Random-Access Memory (SRAM) to improve the attention efficiency.


\section{Methodologies}\label{sec:methodologies}
\begin{figure}[!t]
	\centering
\includegraphics[width=\linewidth]{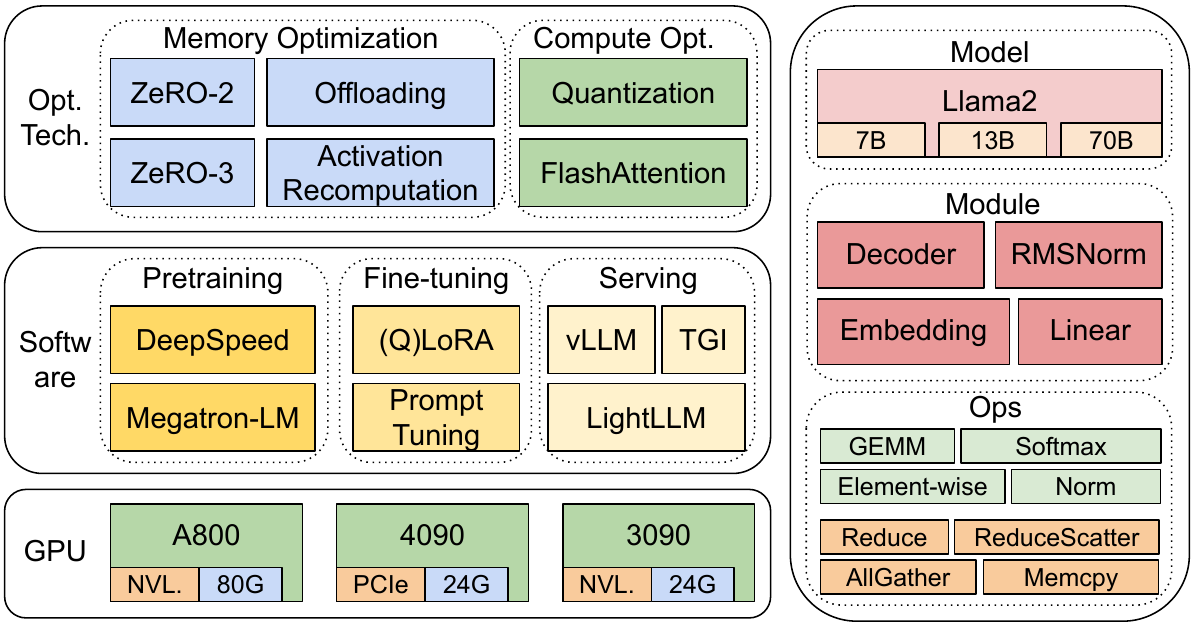}
	\caption{An overview of our benchmarks that cover hardware, software (w/ and w/o optimization techniques), and models.}
	\label{fig:methodoverview}
\end{figure}

Our benchmarks use a top-down methodology that covers the end-to-end step time performance, module-level time performance, and operators time performance of Llama2~\cite{touvron2023llama_2} on three 8-GPU hardware platforms as shown in Figure~\ref{fig:methodoverview}. 

\begin{table}[!t]
	\centering
		\caption{Hardware platform specifications. Each platform has 8x Nvidia GPUs.}
		\label{tab:gpuconfigs}
  \small
		\begin{tabular}{c|c|c|c}
			\hline
			Platform & A800 & RTX4090 & RTX3090 \\\hline
			GPU & A800-80G & RTX4090-24G & RTX3090-24G \\
			    & @ 1.155GHz & @ 2.235GHz & @ 1.395GHz \\\hline
			CPU & 2x AMD  & 2x Intel(R)& 2x AMD \\
                    & $\text{EPYC}^{\text{TM}}$ 7402 &   Xeon(R) Gold 6230 & $\text{EPYC}^{\text{TM}}$ 7302\\
                    & @ 2.80GHz &  @ 2.10GHz & @ 3.00GHz \\\hline
			Memory & 512GiB DDR4 & 512GB DDR4 & 128GB \\\hline
			 Network & NVLink & PCIe4.0x16 & NVLink  \\\hline
		\end{tabular}
\vspace{-15px}
\end{table}
\textbf{Hardware. }In the aspect of hardware evaluation, we cover three 8-GPU platforms, of which the configurations are shown in Table~\ref{tab:gpuconfigs}. In these platforms, we measure the time performance in pre-training, fine-tuning, and serving of LLMs using the different software (e.g., DeepSpeed and Megatron-LM). 

\textbf{Software. }In the aspect of software evaluation, we compare DeepSpeed and Megatron-LM with end-to-end step time in pre-training and fine-tuning. To evaluate optimization techniques, we use DeepSpeed to enable these optimizations (i.e., ZeRO-2, ZeRO-3, offloading, activation recomputation, quantization, and FlashAttention) one by one to measure performance improvement and degradation in time and memory consumption. On LLM serving, there are three highly optimized systems, vLLM~\cite{kwon2023efficient}, LightLLM~\cite{LightLLM2023}, and TGI~\cite{TGI2023}. We compare their performance (latency and throughput) on three testbeds. To understand the end-to-end performance in-depth, we provide microbenchmarks on the performance of model modules and operators that are fundamental components in pre-training, fine-tuning, and inference serving pipelines.

\textbf{Datasets. }In order to ensure the accuracy and reproducibility of the results, we calculated the average length of instruction, input and output of the commonly used dataset alpaca for LLM, i.e., 350 tokens per sample, and randomly generated strings to reach a sequence length of 350, in pre-training, fine-tuning and module-wise analysis. In inference serving, to comprehensively utilize the computational resources and evaluate the robustness and efficiency of the frameworks, all requests were dispatched in a burst pattern. The experimental dataset consists of 1000 synthetic sentences, each containing 512 input tokens, ensuring a consistent evaluation environment. We consistently maintained the ``max generated tokens length" parameter across all experiments on the same GPU platform to guarantee uniformity and comparability of the results.

\subsection{Measuring End-to-end Performance}\label{subsec:endtoend-evaluation}
We measure the end-to-end performance by using the metrics of step time, throughput, and memory consumption in pre-training, fine-tuning, and serving three sizes of Llama2 models (Llama2-7B, Llama2-13B, and Llama2-70B) on three testbeds. 

\textbf{Pre-training (\S\ref{sec:exp-pretraining}).} We first compare the performance (throughput, time, and memory) differences between DeepSpeed and Magetron-LM. Then, we use DeepSpeed to evaluate the impact of optimization techniques ZeRO-2, ZeRO-3, offloading, activation recomputation, quantization, and FlashAttention on both time and memory efficiency on our testbeds.
To understand the root reasons resulting in the measured performance, we further measure the module and operators performance.

\textbf{Fine-tuning (\S\ref{sec:exp-finetuning}).} We compare two popular fine-tuning techniques, LoRA and QLoRA with the baseline (full parameter tuning, or Full-FT) on three testbeds with the metrics of throughput and memory consumption. 

\textbf{Inference Serving (\S\ref{sec:exp-inference}).} We evaluate three widely-recognized inference serving systems: vLLM~\cite{kwon2023efficient}, LightLLM~\cite{LightLLM2023}, and TGI~\cite{TGI2023}, using three testbeds, focusing on metrics such as latency, throughput, and memory consumption. Initially, we deployed the API servers for each of these frameworks. Subsequently, a benchmarking script was used, leveraging \texttt{asyncio}, to dispatch HTTP requests to the model server. As an acknowledged bug\footnote{\href{https://github.com/NVIDIA/nccl-tests/issues/117\#issuecomment-1474324074}{RTX 40x0 NCCL Issue Comment}} exists for the RTX40X0 GPU series, to rectify this issue and ensure that the inference framework functions appropriately on the RTX4090, the configuration \texttt{NCCL\_P2P\_DISABLE=1} was applied. However, this configuration might impact the final performance, putting RTX4090 at a disadvantage against other platforms.

\subsection{Measuring Module-wise Performance}\label{subsec:module-evaluation}
An LLM typically consists of a series of modules (or layers). Taking the Llama2 model as an example. The top level class LlamaForCausalLM consists of one LlamaModel module with a linear layer for downstream tasks. Each module has its own sub-modules which may have unique computation and communication characteristics. Specifically, a LlamaModel is formed by an embedding layer and multiple decoder layers (LlamaDecoderLayer)~\cite{vaswani2017attention}, and the number of LlamaDecoderLayer is configurable. LlamaAttention is the self-attention layer~\cite{vaswani2017attention} that consists of four linear layers for computing $Q$, $K$, $V$, and $O$ projections and one embedding layer (LlamaRotaryEmbedding). LlamaMLP consists of three linear layers whose sizes are configurable and one SiLU activation~\cite{elfwing2018sigmoid} layer (SiLUActivation). LlamaRMSNorm is the RMS normalization~\cite{zhang2019root} layer. In summary, the key modules that form the Llama2 model are Embedding (naive Embedding and LlamaRotaryEmbedding), LlamaDecoderLayer, Linear, SiLUActivation, and LlamaRMSNorm.

In fine-tuning, different approaches introduce extra modules for updating model parameters or adapter parameters. Particularly, LoRA requires extra Linear layers, i.e., the low-rank adapters. QLoRA has a similar training paradigm to LoRA, but its computation uses low-bit representation, which results in low-precision Linear layers, such as 8-bit or 4-bit Linear layers.

\section{Results on Pre-training}\label{sec:exp-pretraining}
In this section, we first analyze the pre-training performance (iteration time or throughput and memory consumption) on different model sizes (7B, 13B, 70B) on three testbeds (\S\ref{subsec:exp-endtoend}), followed by module-wise and ops-level micro-benchmarks (\S\ref{subsec:exp-microbenchmarks}). Unless otherwise specified, each metric (iteration time or throughput and memory consumption) is measured three times for each task and the average is reported.

\subsection{End-to-End Performance} \label{subsec:exp-endtoend}
\begin{table}[]
\centering
\caption{Performance comparison of Megatron and DeepSpeed in pre-training Llama2-7B on the 8-GPU A800-80GB platform. The tensor parallel size of Megatron is 1. We use alpaca dataset and set the sequence length to 350 for both. Any other optimization techniques are excluded.}
\begin{tabular}{l|c|cc}
\hline
Framework & BS & Throughput (Tokens/s) & Memory (GB) \\ \hline
\multirow{2}{*}{Megatron} & 1 & 10936 & 49.1 \\
 & 32 & 13977 & 55.6 \\ \hline
\multirow{2}{*}{DeepSpeed} & 1 & 7488 & 66.76 \\
 & 4 & 19348 & 72.64 \\ \hline
\end{tabular}
\label{table:ds-vs-meg}
\vspace{-15px}
\end{table}

\subsubsection{Megatron-LM vs. DeepSpeed} 
We first conduct an experiment to compare the performance between Megatron-LM and DeepSpeed, neither of which use any memory optimization techniques e.g., ZeRO, in pre-training Llama2-7B on the A800-80GB server. We use a sequence length of 350, and two sets of batch sizes (BS) for both Megatron-LM and DeepSpeed, from 1 to the maximum batch size. We report the training throughput (tokens per second, or tokens/s) and consumed GPU memory in GB as the benchmarks. The results are presented in Table \ref{table:ds-vs-meg}. The results show that \textit{Megatron-LM performs slightly faster than DeepSpeed when the batch size equals 1}, however \textit{DeepSpeed leads the board in terms of training speed when they reach the maximum batch size.} \textit{At the same batch size, DeepSpeed consumes more GPU memory}, compared to the tensor parallel based Megatron-LM. Both systems take up a considerable amount of GPU memory even if the batch size is small, which causes out-of-memory (OOM) on the RTX4090 or RTX3090 GPU servers.

\subsubsection{GPU Scaling Efficiency} 
\begin{figure}[!t]
	\centering
\includegraphics[width=0.8\linewidth]{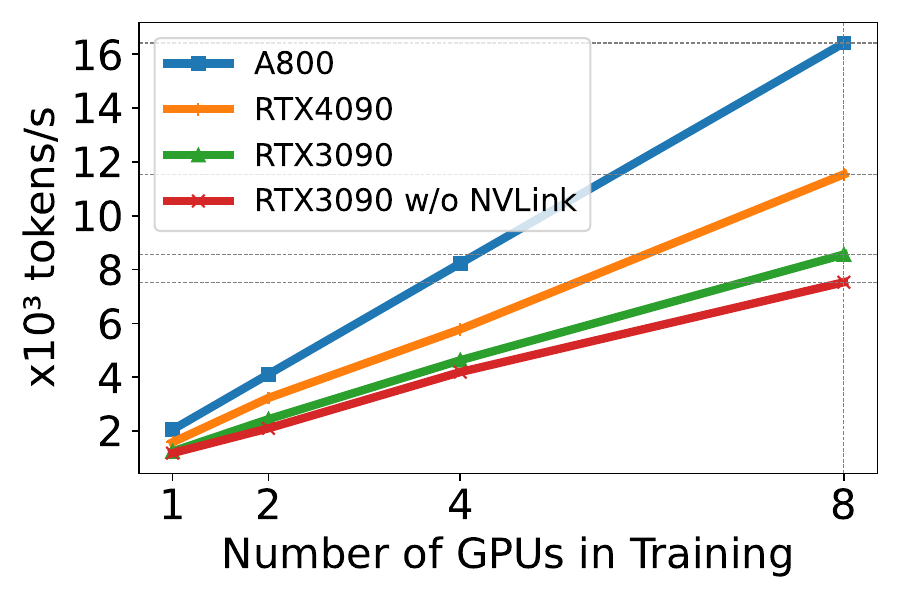}
	\caption{Data parallel training efficiency when training Llama2-7B under different scales of GPUs.}
	\label{fig:scaling}
\end{figure}

We use DeepSpeed with quantization to study scaling efficiency (from 1 GPU to 8 GPUs) on different hardware platforms in training Llama2-7B (sequence length is 350, batch size is 2). The results are presented in Figure \ref{fig:scaling}, where the slope indicates scaling efficiency. The figure shows that A800 has almost linear scaling, whereas RTX4090 and RTX3090 have slightly low scaling efficiency (90.8\% and 85.9\% respectively). RTX4090 achieves 4.9\% higher scaling efficiency than RTX3090. In the RTX3090 platform, NVLink connection helps improve the scaling efficiency by 10\% over without NVLink.

\subsubsection{Hardware and Optimized Techniques} We use DeepSpeed to evaluate the training performance under different memory- and compute-efficient methods. We use a sequence length of 350, and batch size of 1 across all evaluations for a fair comparison, and load the model weight into bf16 by default for all experiments. For ZeRO-2 and ZeRO-3 with offloading, we offload the optimizer state and optimizer state + model weights to CPU RAM, respectively. For quantization, we use the configuration of 4 bits with double quantization as suggested in a prior study\cite{dettmers2023qlora}. We also report the performance on RTX3090 when NVLink is disabled, i.e., all data are transmitted by PCIe buses. The results are presented in Table \ref{apx-table:many-method-bs1}.

\begin{table*}[]
\centering
\caption{Pre-training performance comparison among the baseline setting (Naive), ZeRO-2 (Z2) \& 3 (Z3), offloading (O), 4-bit quantization (Q), activation recomputation (R), and FlashAttention (F) on three types of 8-GPU platforms. In offloading, ZeRO-3 offloads both optimizer state and parameters to CPU, while ZeRO-2 only offloads optimizer state to CPU. We set the batch size to 1, sequence length to 350 for all results. We report tokens/s (Tokens/s) as the throughput, average with three independent runs with standard deviation shown in the bottom-right corner of each throughput value and peak GPU memory usage (M) in GB. In each run, throughput is averaged over 100 steps after 30 warm-up steps. "-" indicates OOM.}
\addtolength{\tabcolsep}{-4pt}
\begin{tabular}{l|l|cccccccc}
\hline
\multirow{3}{*}{Model} & \multirow{3}{*}{Method} & \multicolumn{8}{c}{Hardware platform} \\ \cline{3-10} 
 &  & \multicolumn{2}{c|}{A800} & \multicolumn{2}{c|}{RTX4090} & \multicolumn{2}{c|}{RTX3090 w/ NVLink} & \multicolumn{2}{c}{RTX3090 w/o NVLink} \\ \cline{3-10} 
 &  & Tokens/s & \multicolumn{1}{c|}{M (GB)} & Tokens/s & \multicolumn{1}{c|}{M (GB)} & Tokens/s & \multicolumn{1}{c|}{M (GB)} & Tokens/s & M (GB) \\ \hline
\multirow{22}{*}{7B} & Naive & $7488.3_{.07}$ & \multicolumn{1}{c|}{66.7} & \multicolumn{2}{c|}{-} & \multicolumn{2}{c|}{-} & \multicolumn{2}{c}{-} \\
 & Z2 & $6101.6_{.03}$ & \multicolumn{1}{c|}{37.8} & \multicolumn{2}{c|}{-} & \multicolumn{2}{c|}{-} & \multicolumn{2}{c}{-} \\
 & Z2+O & $393.9_{.0}$ & \multicolumn{1}{c|}{32.8} & $67.7_{.0}$ & \multicolumn{1}{c|}{19.1} & $58._{.01}$ & \multicolumn{1}{c|}{19} & $50.5_{.01}$ & 19 \\
 & Z3 & $5491.4_{.02}$ & \multicolumn{1}{c|}{30.5} & $129.3_{.03}$ & \multicolumn{1}{c|}{22.6} & $90.8_{.0}$ & \multicolumn{1}{c|}{22.6} & $82.9_{.01}$ & 22.6 \\
 & Z3+O & $271.8_{.0}$ & \multicolumn{1}{c|}{10.4} & $64.4_{.0}$ & \multicolumn{1}{c|}{10.4} & $48.8_{.02}$ & \multicolumn{1}{c|}{10.4} & $39.9_{.02}$ & 10.4 \\
 & Q & $10813.4_{.01}$ & \multicolumn{1}{c|}{9.8} & $4879.2_{.05}$ & \multicolumn{1}{c|}{10.1} & $3424.4_{.02}$ & \multicolumn{1}{c|}{9.8} & $2916.5_{0.02}$ & 9.8 \\
 & R & $7236.8_{.12}$ & \multicolumn{1}{c|}{65.9} & \multicolumn{2}{c|}{-} & \multicolumn{2}{c|}{-} & \multicolumn{2}{c}{-} \\
 & F & $7694.1_{.03}$ & \multicolumn{1}{c|}{66.7} & \multicolumn{2}{c|}{-} & \multicolumn{2}{c|}{-} & \multicolumn{2}{c}{-} \\
 & R+Z2 & $5704._{.04}$ & \multicolumn{1}{c|}{38.1} & \multicolumn{2}{c|}{-} & \multicolumn{2}{c|}{-} & \multicolumn{2}{c}{-} \\
 & R+Z2+O & $402.7_{.0}$ & \multicolumn{1}{c|}{29.6} & $74.1_{.02}$ & \multicolumn{1}{c|}{19} & $44.1_{.03}$ & \multicolumn{1}{c|}{19} & $46.1_{.03}$ & 19 \\
 & R+Z3 & $4738.8_{.02}$ & \multicolumn{1}{c|}{28.8} & $127.5_{.03}$ & \multicolumn{1}{c|}{22.6} & $85.8_{.0}$ & \multicolumn{1}{c|}{22.6} & $71.7_{.0}$ & 22.6 \\
 & R+Z3+O & $266.7_{.0}$ & \multicolumn{1}{c|}{6.4} & $65.2_{.0}$ & \multicolumn{1}{c|}{6.4} & $45.1_{.01}$ & \multicolumn{1}{c|}{6.4} & $38.1_{.0}$ & 6.4 \\
 & R+Q & $7126.4_{.05}$ & \multicolumn{1}{c|}{6} & $4699._{.05}$ & \multicolumn{1}{c|}{6} & $2377.2_{.06}$ & \multicolumn{1}{c|}{6} & $2120.5_{.04}$ & 6.0 \\
 & R+F & $7528.7_{.05}$ & \multicolumn{1}{c|}{66.1} & \multicolumn{2}{c|}{-} & \multicolumn{2}{c|}{-} & \multicolumn{2}{c}{-} \\
 & F+Z2 & $6322._{.03}$ & \multicolumn{1}{c|}{38.2} & \multicolumn{2}{c|}{-} & \multicolumn{2}{c|}{-} & \multicolumn{2}{c}{-} \\
 & F+Z2+O & $403.2_{.06}$ & \multicolumn{1}{c|}{32} & $78.2_{.0}$ & \multicolumn{1}{c|}{18.1} & $56.6_{.01}$ & \multicolumn{1}{c|}{18} & $51._{.0}$ & 18 \\
 & F+Z3 & $5590.1_{.05}$ & \multicolumn{1}{c|}{29.2} & $154.2_{.03}$ & \multicolumn{1}{c|}{21.6} & $97.6_{.0}$ & \multicolumn{1}{c|}{21.4} & $82.6_{.01}$ & 21.4 \\
 & F+Z3+O & $272.8_{.01}$ & \multicolumn{1}{c|}{8.8} & $66.5_{.0}$ & \multicolumn{1}{c|}{8.8} & $49.5_{.0}$ & \multicolumn{1}{c|}{8.8} & $38.7_{.02}$ & 8.8 \\
 & F+R+Z2 & $5984.3_{.05}$ & \multicolumn{1}{c|}{38.1} & \multicolumn{2}{c|}{-} & \multicolumn{2}{c|}{-} & \multicolumn{2}{c}{-} \\
 & F+R+Z2+O & $402.2_{.01}$ & \multicolumn{1}{c|}{29.6} & $74.4_{.05}$ & \multicolumn{1}{c|}{17.7} & $50.1_{.04}$ & \multicolumn{1}{c|}{17.7} & $49.6_{.2}$ & 17.7 \\
 & F+R+Z3 & $4803.8_{.12}$ & \multicolumn{1}{c|}{27.4} & $130.8_{.03}$ & \multicolumn{1}{c|}{21} & $94.4_{.0}$ & \multicolumn{1}{c|}{21} & $82._{.01}$ & 21 \\
 & F+R+Z3+O & $270._{.01}$ & \multicolumn{1}{c|}{6.7} & $61.8_{.01}$ & \multicolumn{1}{c|}{6.7} & $47._{.0}$ & \multicolumn{1}{c|}{6.5} & $44.8_{.0}$ & 6.5 \\ \hline
\multirow{16}{*}{13B} & Z2 & $3234._{.06}$ & \multicolumn{1}{c|}{71.4} & \multicolumn{2}{c|}{-} & \multicolumn{2}{c|}{-} & \multicolumn{2}{c}{-} \\
 & Z2+O & $196.2_{.0}$ & \multicolumn{1}{c|}{57.9} & \multicolumn{2}{c|}{-} & \multicolumn{2}{c|}{-} & \multicolumn{2}{c}{-} \\
 & Z3 & $3670.5_{.02}$ & \multicolumn{1}{c|}{48.9} & \multicolumn{2}{c|}{-} & \multicolumn{2}{c|}{-} & \multicolumn{2}{c}{-} \\
 & Z3+O & $132.8_{.01}$ & \multicolumn{1}{c|}{12.7} & $23.8_{.01}$ & \multicolumn{1}{c|}{12.7} & $18.1_{.0}$ & \multicolumn{1}{c|}{12.2} & $16.6_{.0}$ & 12.2 \\
 & R+Z2 & $3064.1_{.05}$ & \multicolumn{1}{c|}{71.8} & \multicolumn{2}{c|}{-} & \multicolumn{2}{c|}{-} & \multicolumn{2}{c}{-} \\
 & R+Z2+O & $198.9_{.0}$ & \multicolumn{1}{c|}{53.1} & \multicolumn{2}{c|}{-} & \multicolumn{2}{c|}{-} & \multicolumn{2}{c}{-} \\
 & R+Z3 & $3318.2_{.03}$ & \multicolumn{1}{c|}{48.9} & \multicolumn{2}{c|}{-} & \multicolumn{2}{c|}{-} & \multicolumn{2}{c}{-} \\
 & R+Z3+O & $130.9_{.02}$ & \multicolumn{1}{c|}{7.8} & $22.3_{.01}$ & \multicolumn{1}{c|}{7.8} & $17.2_{.0}$ & \multicolumn{1}{c|}{7.8} & $15.5_{.0}$ & 7.8 \\
 & F+Z2 & $3275.6_{.04}$ & \multicolumn{1}{c|}{72.2} & \multicolumn{2}{c|}{-} & \multicolumn{2}{c|}{-} & \multicolumn{2}{c}{-} \\
 & F+Z2+O & $198.6_{.02}$ & \multicolumn{1}{c|}{56.8} & \multicolumn{2}{c|}{-} & \multicolumn{2}{c|}{-} & \multicolumn{2}{c}{-} \\
 & F+Z3 & $3680.2_{.07}$ & \multicolumn{1}{c|}{52.2} & \multicolumn{2}{c|}{-} & \multicolumn{2}{c|}{-} & \multicolumn{2}{c}{} \\
 & F+Z3+O & $134.2_{.03}$ & \multicolumn{1}{c|}{11.5} & $32.3_{.02}$ & \multicolumn{1}{c|}{11.5} & $19.4_{.0}$ & \multicolumn{1}{c|}{11.3} & $17._{.0}$ & 11.3 \\
 & F+R+Z2 & $3900.5_{.05}$ & \multicolumn{1}{c|}{71.7} & \multicolumn{2}{c|}{-} & \multicolumn{2}{c|}{-} & \multicolumn{2}{c}{-} \\
 & F+R+Z2+O & $202._{.01}$ & \multicolumn{1}{c|}{52.9} & \multicolumn{2}{c|}{-} & \multicolumn{2}{c|}{-} & \multicolumn{2}{c}{-} \\
 & F+R+Z3 & $3483.4_{.06}$ & \multicolumn{1}{c|}{53.7} & \multicolumn{2}{c|}{-} & \multicolumn{2}{c|}{-} & \multicolumn{2}{c}{-} \\
 & F+R+Z3+O & $134._{.01}$ & \multicolumn{1}{c|}{7.9} & $22.3_{.0}$ & \multicolumn{1}{c|}{7.9} & $17.4_{.0}$ & \multicolumn{1}{c|}{7.9} & $15.9_{.0}$ & 7.9 \\ \hline
\end{tabular}
\label{apx-table:many-method-bs1}
\end{table*}


\textbf{Hardware Impact.} (1) The throughput of A800 exceeds 50 times that of RTX4090 and RTX3090 GPUs on all evaluated cases except quantization. In the case of using quantization, RTX GPUs can achieve half of the A800 performance. (2) RTX4090 is 50\% better than RTX3090, and NVLink in RTX3090 helps improve the performance by around 10\%. (3) Since A800 has 80GB memory, while RTX4090 and RTX3090 have only 24GB each, so some cases, e.g., Naive and ZeRO-2 cannot run on RTX4090 and RTX3090 GPUs. (4) With ZeRO (and offloading), a 8x 80GB (8x 24GB) GPU server can at most fit a 30B model for mixed half-precision training. 

\textbf{Optimization Techniques.} In pre-training Llama2-7B, ZeRO-2's GPU memory consumption is about 57\% of Naive's, without sacrificing model performance and training efficiency. Meanwhile, ZeRO-3 performs slightly slower than ZeRO-2, with less memory consumption. However, ZeRO-3 outperforms ZeRO-2 when pre-training Llama2-13B. This difference is because sharding the full model state helps reduce communication further, especially when training larger models.
Offloading significantly slows down the training process as it offloads some shards and computing to RAM and CPU, and reduce the GPU memory consumption. Quantization achieves the highest throughput on all hardware platforms, but may affect convergence. FlashAttention also accelerates training and can be used together with memory-efficient methods, such as ZeRO. Activation recomputation further reduces GPU memory usage but decreases throughput. Note that the activation memory is small when batch size = 1, and activation recomputation could save more memory with higher batch sizes. In Table \ref{apx-table:many-method-bs1}, when using ZeRO or offloading, memory consumption of the same method varies across platforms. Specifically, it takes more memory on A800 than on the other platforms. This difference is because memory are pinned on CPU in sharding and offloading, and the handles are dynamically loaded into GPU memory, based on available physical memory which is larger on A800 than the other platforms. This table also demonstrates that training Llama2-13B achieves half of the Llama2-7B training's throughput. With such model performance between Llama2-7B and Llama2-13B, training a model with 13B parameters may be a better choice than a 7B model. We further leverage the computing power of different GPU servers, by maximizing the batch sizes of each method to get the maximum throughput. The result is presented in Table \ref{apx-table:many-method-bs32}.

In this table, we find that when batch size is 16, FlashAttention with ZeRO-3 and offloading consumes even more GPU memory compared to FlashAttention with ZeRO3 (77.5GB vs. 73GB). When conducting this experiment, ZeRO-3 with offloading often yields an inbalanced device map, i.e, it takes more memory on GPU 0 than on other GPUs. We will address this problem in future studies. Overall, Table \ref{apx-table:many-method-bs32} shows that enlarging the batch size easily boosts the training process, which also overlaps communication and GPU computing. Hence, a GPU server with high bandwidth and large GPU memory is more suitable for full parameter mix-precision pre-training than a consumer-level GPU server.

\begin{table*}[!ht]
\centering
\caption{Pre-training performance comparison among 6 methods when maximizing the batch size to get the maximum throughput. We use the same configuration as Table \ref{apx-table:many-method-bs1}.}
\addtolength{\tabcolsep}{-5pt}
\begin{tabular}{l|l|cclcclcclccl}
\hline
\multirow{3}{*}{Model} & \multirow{3}{*}{Method} & \multicolumn{12}{c}{Hardware platform} \\ \cline{3-14} 
 &  & \multicolumn{3}{c}{A800} & \multicolumn{3}{c}{RTX4090} & \multicolumn{3}{c}{RTX3090 w/ NVLink} & \multicolumn{3}{c}{RTX3090 w/o NVLink} \\ \cline{3-14} 
 &  & Tokens/s & M (GB) & \multicolumn{1}{l|}{BS} & Tokens/s & M (GB) & \multicolumn{1}{l|}{BS} & Tokens/s & M (GB) & \multicolumn{1}{l|}{BS} & Tokens/s & M (GB) & BS \\ \hline
\multirow{22}{*}{7B} & Naive & $19348.3_{.05}$ & 72.6 & \multicolumn{1}{l|}{4} & \multicolumn{3}{c|}{-} & \multicolumn{3}{c|}{-} & \multicolumn{3}{c}{-} \\
 & Z2 & $21481.7_{.04}$ & 58.0 & \multicolumn{1}{l|}{8} & \multicolumn{3}{c|}{-} & \multicolumn{3}{c|}{-} & \multicolumn{3}{c}{-} \\
 & Z2+O & $5196.6_{.02}$ & 77.7 & \multicolumn{1}{l|}{16} & $146.1_{.0}$ & 22.5 & \multicolumn{1}{l|}{2} & $110._{.0}$ & 22.5 & \multicolumn{1}{l|}{2} & $106.5_{.0}$ & 22.5 & 2 \\
 & Z3 & $16789.4_{.05}$ & 77.2 & \multicolumn{1}{l|}{16} & $240.4_{.01}$ & 22.6 & \multicolumn{1}{l|}{2} & $184._{.01}$ & 22.68 & \multicolumn{1}{l|}{2} & $172._{.0}$ & 22.6 & 2 \\
 & Z3+O & $4066.3_{.02}$ & 74.9 & \multicolumn{1}{l|}{16} & $215.8_{.02}$ & 20.5 & \multicolumn{1}{l|}{4} & $161.5_{.01}$ & 20.5 & \multicolumn{1}{l|}{4} & $128.4_{.02}$ & 20.5 & 4 \\
 & Q & $27902.1_{.05}$ & 49.8 & \multicolumn{1}{l|}{8} & $13899.1_{.07}$ & 22.7 & \multicolumn{1}{l|}{4} & $10790._{.05}$ & 22.7 & \multicolumn{1}{l|}{4} & $8507.9_{.03}$ & 22.7 & 4 \\
 & R & $22909.4_{.08}$ & 75.1 & \multicolumn{1}{l|}{64} & \multicolumn{3}{c|}{-} & \multicolumn{3}{c|}{-} & \multicolumn{3}{c}{-} \\
 & F & $25872.3_{.07}$ & 76.4 & \multicolumn{1}{l|}{8} & \multicolumn{3}{c|}{-} & \multicolumn{3}{c|}{-} & \multicolumn{3}{c}{-} \\
 & R+Z2 & $22608.8_{.02}$ & 45.9 & \multicolumn{1}{l|}{64} & \multicolumn{3}{c|}{-} & \multicolumn{3}{c|}{-} & \multicolumn{3}{c}{-} \\
 & R+Z2+O & $12322.4_{.3}$ & 58.5 & \multicolumn{1}{l|}{64} & $1135.3_{.02}$ & 19.8 & \multicolumn{1}{l|}{16} & $891.2_{.02}$ & 19.8 & \multicolumn{1}{l|}{16} & $877.8_{.04}$ & 19.8 & 16 \\
 & R+Z3 & $22691.8_{.05}$ & 59.2 & \multicolumn{1}{l|}{64} & $2033.7_{.05}$ & 22.8 & \multicolumn{1}{l|}{16} & $1244.7_{.05}$ & 22.8 & \multicolumn{1}{l|}{16} & $1172.7_{.01}$ & 22.8 & 16 \\
 & R+Z3+O & $10252.9_{.03}$ & 52.8 & \multicolumn{1}{l|}{64} & $3595.4_{.02}$ & 22.5 & \multicolumn{1}{l|}{64} & $1444.8_{.01}$ & 17.5 & \multicolumn{1}{l|}{32} & $1241.3_{.02}$ & 17.5 & 32 \\
 & R+Q & $24078.8_{.02}$ & 56.8 & \multicolumn{1}{l|}{64} & $14786.5_{.12}$ & 20 & \multicolumn{1}{l|}{32} & $7563.9_{.1}$ & 18.9 & \multicolumn{1}{l|}{16} & $5992.2_{.07}$ & 18.9 & 16 \\
 & R+F & $25639.1_{.05}$ & 73.1 & \multicolumn{1}{l|}{64} & \multicolumn{3}{c|}{-} & \multicolumn{3}{c|}{-} & \multicolumn{3}{c}{-} \\
 & F+Z2 & $27340.3_{.11}$ & 76.5 & \multicolumn{1}{l|}{16} & \multicolumn{3}{c|}{-} & \multicolumn{3}{c|}{-} & \multicolumn{3}{c}{-} \\
 & F+Z2+O & $5496.2_{.12}$ & 70.3 & \multicolumn{1}{l|}{16} & $78.2_{.0}$ & 18.1 & \multicolumn{1}{l|}{1} & $56.6_{.01}$ & 18 & \multicolumn{1}{l|}{1} & $51._{.0}$ & 18 & 1 \\
 & F+Z3 & $27936.9_{.1}$ & 73 & \multicolumn{1}{l|}{16} & $154.2_{.01}$ & 21.6 & \multicolumn{1}{l|}{1} & $97.6_{.0}$ & 21.4 & \multicolumn{1}{l|}{1} & $82.6_{.01}$ & 21.4 & 1 \\
 & F+Z3+O & $3651.4_{.15}$ & 77.5 & \multicolumn{1}{l|}{16} & $1877.4_{.03}$ & 19.4 & \multicolumn{1}{l|}{16} & $1443.1_{.05}$ & 19.4 & \multicolumn{1}{l|}{16} & $1115.1_{.04}$ & 19.4 & 16 \\
 & F+R+Z2 & $25270.5_{.06}$ & 45.4 & \multicolumn{1}{l|}{64} & \multicolumn{3}{c|}{-} & \multicolumn{3}{c|}{-} & \multicolumn{3}{c}{-} \\
 & F+R+Z2+O & $12856.3_{.1}$ & 61 & \multicolumn{1}{l|}{64} & $74.4_{.0}$ & 17.7 & \multicolumn{1}{l|}{1} & $50.1_{.04}$ & 17.7 & \multicolumn{1}{l|}{1} & $49.6_{.2}$ & 17.7 & 1 \\
 & F+R+Z3 & $25160.5_{.22}$ & 56.7 & \multicolumn{1}{l|}{64} & $130.8_{.01}$ & 21 & \multicolumn{1}{l|}{1} & $94.4_{.0}$ & 21 & \multicolumn{1}{l|}{1} & $82_{.01}$ & 21 & 1 \\
 & F+R+Z3+O & $10801._{.15}$ & 49.8 & \multicolumn{1}{l|}{64} & $2009._{.01}$ & 17.2 & \multicolumn{1}{l|}{32} & $1455.1_{.01}$ & 17.2 & \multicolumn{1}{l|}{32} & $1243.8_{.01}$ & 17.2 & 32 \\ \hline
\multirow{16}{*}{13B} & Z2 & $8610.8_{.07}$ & 71.8 & \multicolumn{1}{l|}{4} & \multicolumn{3}{c|}{-} & \multicolumn{3}{c|}{-} & \multicolumn{3}{c}{-} \\
 & Z2+O & $1106.3_{.03}$ & 78.9 & \multicolumn{1}{l|}{8} & \multicolumn{3}{c|}{-} & \multicolumn{3}{c|}{-} & \multicolumn{3}{c}{-} \\
 & Z3 & $8891._{.05}$ & 78.5 & \multicolumn{1}{l|}{8} & \multicolumn{3}{c|}{-} & \multicolumn{3}{c|}{-} & \multicolumn{3}{c}{-} \\
 & Z3+O & $1047.6_{.02}$ & 66.7 & \multicolumn{1}{l|}{8} & $55.8_{.0}$ & 18.8 & \multicolumn{1}{l|}{2} & $47.2_{.0}$ & 18.8 & \multicolumn{1}{l|}{2} & $43.4_{.0}$ & 18.8 & 2 \\
 & R+Z2 & $12503.2_{.05}$ & 75.7 & \multicolumn{1}{l|}{64} & \multicolumn{3}{c|}{-} & \multicolumn{3}{c|}{-} & \multicolumn{3}{c}{-} \\
 & R+Z2+O & $640435_{.13}$ & 77.7 & \multicolumn{1}{l|}{64} & \multicolumn{3}{c|}{-} & \multicolumn{3}{c|}{-} & \multicolumn{3}{c}{-} \\
 & R+Z3 & $12603.2_{.05}$ & 77.8 & \multicolumn{1}{l|}{64} & \multicolumn{3}{c|}{-} & \multicolumn{3}{c|}{-} & \multicolumn{3}{c}{-} \\
 & R+Z3+O & $5344.6_{.2}$ & 63.6 & \multicolumn{1}{l|}{64} & $53.3_{.0}$ & 22 & \multicolumn{1}{l|}{16} & $48.2_{.0}$ & 22 & \multicolumn{1}{l|}{16} & $47.5_{.0}$ & 22 & 16 \\
 & F+Z2 & $8933.5_{.07}$ & 75.7 & \multicolumn{1}{l|}{4} & \multicolumn{3}{c|}{-} & \multicolumn{3}{c|}{-} & \multicolumn{3}{c}{-} \\
 & F+Z2+O & $1263.4_{.12}$ & 27.5 & \multicolumn{1}{l|}{8} & \multicolumn{3}{c|}{-} & \multicolumn{3}{c|}{-} & \multicolumn{3}{c}{-} \\
 & F+Z3 & $14243.7_{.04}$ & 74.4 & \multicolumn{1}{l|}{8} & \multicolumn{3}{c|}{-} & \multicolumn{3}{c|}{-} & \multicolumn{3}{c}{-} \\
 & F+Z3+O & $1999.3_{.09}$ & 74.9 & \multicolumn{1}{l|}{16} & $59.4_{.0}$ & 22.6 & \multicolumn{1}{l|}{4} & $49.9_{.0}$ & 22.6 & \multicolumn{1}{l|}{4} & $47.9_{.0}$ & 22.6 & 4 \\
 & F+R+Z2 & $14862.6_{.04}$ & 71.8 & \multicolumn{1}{l|}{64} & \multicolumn{3}{c|}{-} & \multicolumn{3}{c|}{-} & \multicolumn{3}{c}{-} \\
 & F+R+Z2+O & $7023.3_{.07}$ & 57.9 & \multicolumn{1}{l|}{64} & \multicolumn{3}{c|}{-} & \multicolumn{3}{c|}{-} & \multicolumn{3}{c}{-} \\
 & F+R+Z3 & $14761._{.09}$ & 77.1 & \multicolumn{1}{l|}{64} & \multicolumn{3}{c|}{-} & \multicolumn{3}{c|}{-} & \multicolumn{3}{c}{-} \\
 & F+R+Z3+O & $5749.7_{.09}$ & 34.9 & \multicolumn{1}{l|}{64} & $173.4_{.0}$ & 23 & \multicolumn{1}{c|}{32} & $167.9_{.0}$ & 23 & \multicolumn{1}{c|}{32} & $165.2_{.0}$ & 23 & \multicolumn{1}{c}{32} \\ \cline{2-14} 
\end{tabular}
\label{apx-table:many-method-bs32}
\end{table*}


\subsection{Module-wise Analysis}\label{subsec:exp-modulewise}

To gain an insight into pre-training performance, we conducted detailed module-wise profiling on the pre-training process. Specifically, we selected the Llama2-7B model on the A800 platform to ensure that all cases can run for performance analysis. The traces were generated using ``torch.profiler'', and all performance numbers presented in this section are averaged over ten steps. The experimental setup aligns with that described in Section \ref{subsec:exp-endtoend}.

\begin{table}[!ht]
\centering
\caption{The overall and compute kernel time cost of forward, backward, and optimizer of one step pre-training Llama2-7B model. *For backward phase, since the overall time includes non-overlapped time, the percentage of compute kernel time is significantly lower than forward phase and optimizer. If the non-overlapped time is removed from backward phase, this value becomes to 94.8.}
\begin{tabular}{c|ccc}
\hline
 & Forward & Backward & Optimizer \\ \hline
Overall (ms)                                                           & 75.0    & 250.0    & 193.9    \\
Breakdown in one step(\%)                                                     & 14.3    & 47.5     & 36.9     \\
\begin{tabular}[c]{@{}c@{}}Compute kernel (ms)\end{tabular} & 68.8   & 200.9   & 181.4  \\ \hline
\end{tabular}
\label{table:microproportion}
\end{table}


Given the memory capacity of the A800, we set the batch size to 2. The time consumed in the forward, backward, and optimizer phases in one pre-training step are detailed in Table \ref{table:microproportion}. Notably, about 37\% of the time is dedicated to the optimizer, which deviates from expectations as the optimizer has only element-wise operations. We analyze this phenomenon in Section \ref{subsec:exp-recompflash}, with a focus on the impact of recomputation.

We conducted module-wise time analysis for both the forward and backward phases, and the results are presented in Table \ref{table:microproportion}.

In Llama2, the decoder layer, built on a Transformer-decoder architecture, accounts for the majority of the computation time. Specifically, the multi-layer perceptron (MLP) and the query, key, and value (QKV) projections, which rely on general matrix multiplications (GEMM) operations, are the most time-consuming components.
Additionally, the RMSNorm and RoPE modules take significant amounts of time due to the great number of element-wise operations. Compared to the forward phase, the backward phase incurs additional communication overhead for gradient synchronization across GPUs. 


\begin{table*}[!ht]
\centering
\caption{The module-wise time consumption and percentage in forward and backward phases of Llama2-7B. The time consumption of modules in the decoder layer is the accumulated time over 32 iterations.}
\begin{tabular}{c|c|c|c|c|c|c|c}
\hline
& Module& Time(ms) & Percentage(\%) & & Module & Time(ms) & Percentage(\%) \\ \hline
\multirow{11}{*}{Forward} & Embedding   & 0.032    & 0.04  & \multirow{11}{*}{Backward} & Embedding & 0.252&0.1 \\ \cline{2-4} \cline{6-8}
& QKV& 9.92     &13.2 &  & QKV & 36.26    & 14.5        \\ 
& RoPE & 6.66  & 8.9&  & RoPE & 15.58    &  6.2       \\  
 & Bmm0  & 4.32  & 5.8& & Bmm0  & 5.63&  2.3\\ 
 & Softmax & 2.62  & 3.5&& Softmax& 4.29 & 1.7  \\ 
& Bmm1  & 2.21     & 2.9 & & Bmm1 & 6.14 &  2.5 \\ 
& Output & 3.39 &   4.5 &  & Output  & 12.32    & 4.9 \\ 
& MLP & 29.06  &  38.7 &  & MLP& 88.70 & 35.5  \\ \cline{2-4} \cline{6-8}
& RMSNorm     & 6.91     & 9.2 &  & RMSNorm & 27.40  &  11.0\\\cline{2-4} \cline{6-8}
& Linear & 1.08 & 1.4 &   & Linear &  2.898 & 1.2  \\ \cline{2-4} \cline{6-8}
 & -    & - &  -&                     &\begin{tabular}[c]{@{}l@{}}non-overlapping\\ communication\end{tabular} &  38.76& 15.5 \\ \hline
\end{tabular}
\label{label:micromoduletime}
\vspace{-15px}
\end{table*}



\subsection{Impact of Recomputation and FlashAttention}\label{subsec:exp-recompflash}
Techniques to accelerate pre-training can be roughly divided into two categories: saving memory to increase batch size and accelerating compute kernels. As shown in Table 5, the GPU is idle 5-10\% of time in forward, backward, and optimizer phases. We believe this idle time is due to the small batch size. We test the maximum batch size that can be used with all available techniques and find that recomputation can increase the batch size from 2 to 32 at its largest. We therefore select recomputation for increasing batch size and flashattention for accelerating compute kernel analysis.


\textbf{Recomputation.} 
As the batch size increases, the time for forward and backward phases increases significantly, with little GPU idle time (Table \ref{table:bs32proportion}). The optimizer updates the model parameters based on optimizer states, hence this process will have a lot of element-wise operations, and the time consumption will remain unchanged in spite of the increase of batch size. In contrast, there are many batch operations in forward and backward phases, which will increase the time consumption with the increase of batch size. Therefore, when the batch size is relatively small, the percentage of time taken by the optimizer will be relatively large; when using the recomputation technique to pre-train with a large batch size, the percentage of time taken by the optimizer will be very small. 

\begin{figure}[!ht]
     \centering
     \begin{subfigure}[b]{0.18\textwidth}
        \centering
        \includegraphics[width=\linewidth]{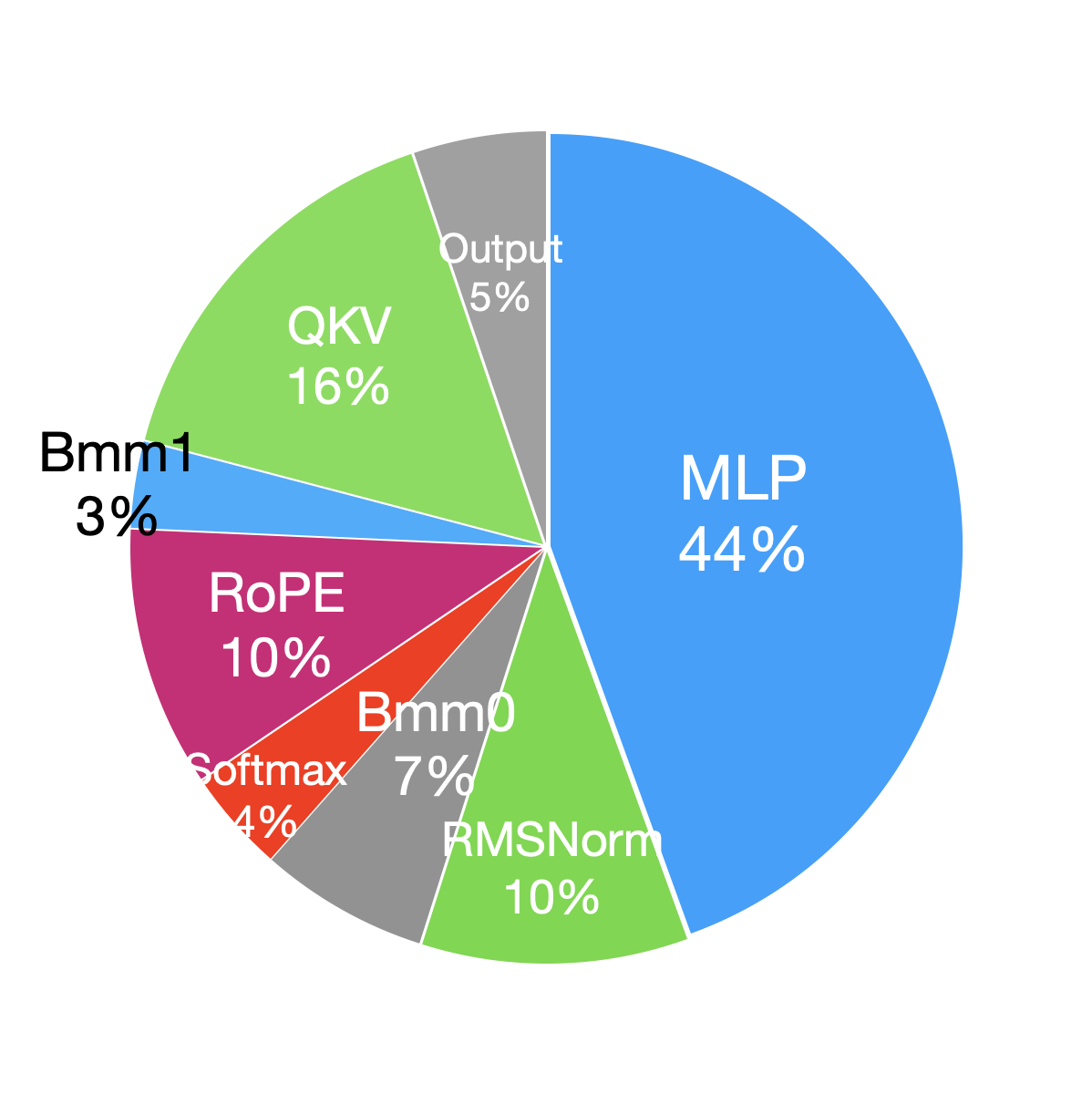}
        \captionsetup{justification=centering}
        \caption*{(a) Forward w/o recomputation}
     \end{subfigure}
    \begin{subfigure}[b]{0.18\textwidth}
        \centering
        \includegraphics[width=\linewidth]{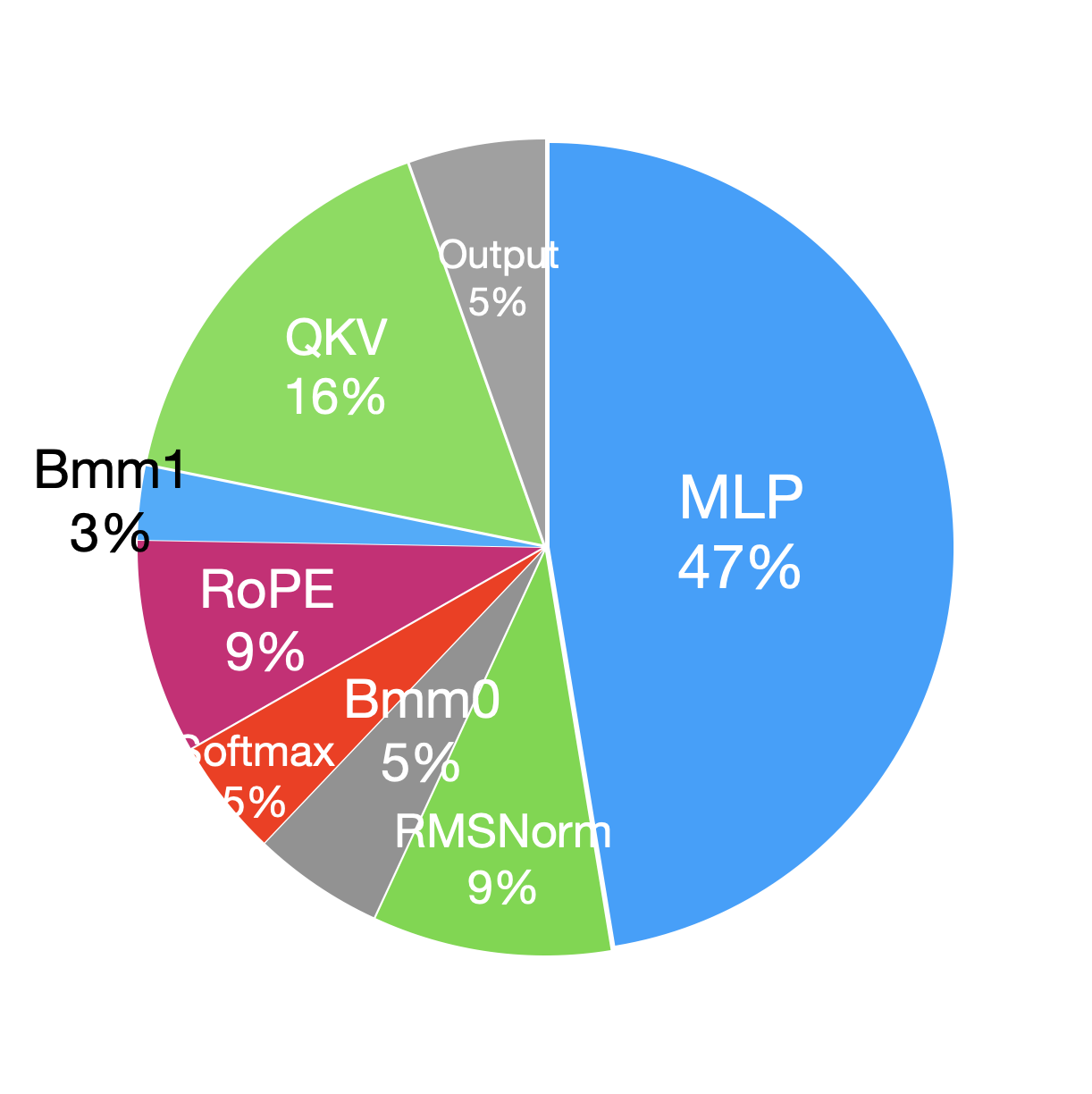}
        \captionsetup{justification=centering}
        \caption*{(b) Forward w/ recomputation}
     \end{subfigure}
     \begin{subfigure}[b]{0.18\textwidth}
        \centering
        \includegraphics[width=\linewidth]{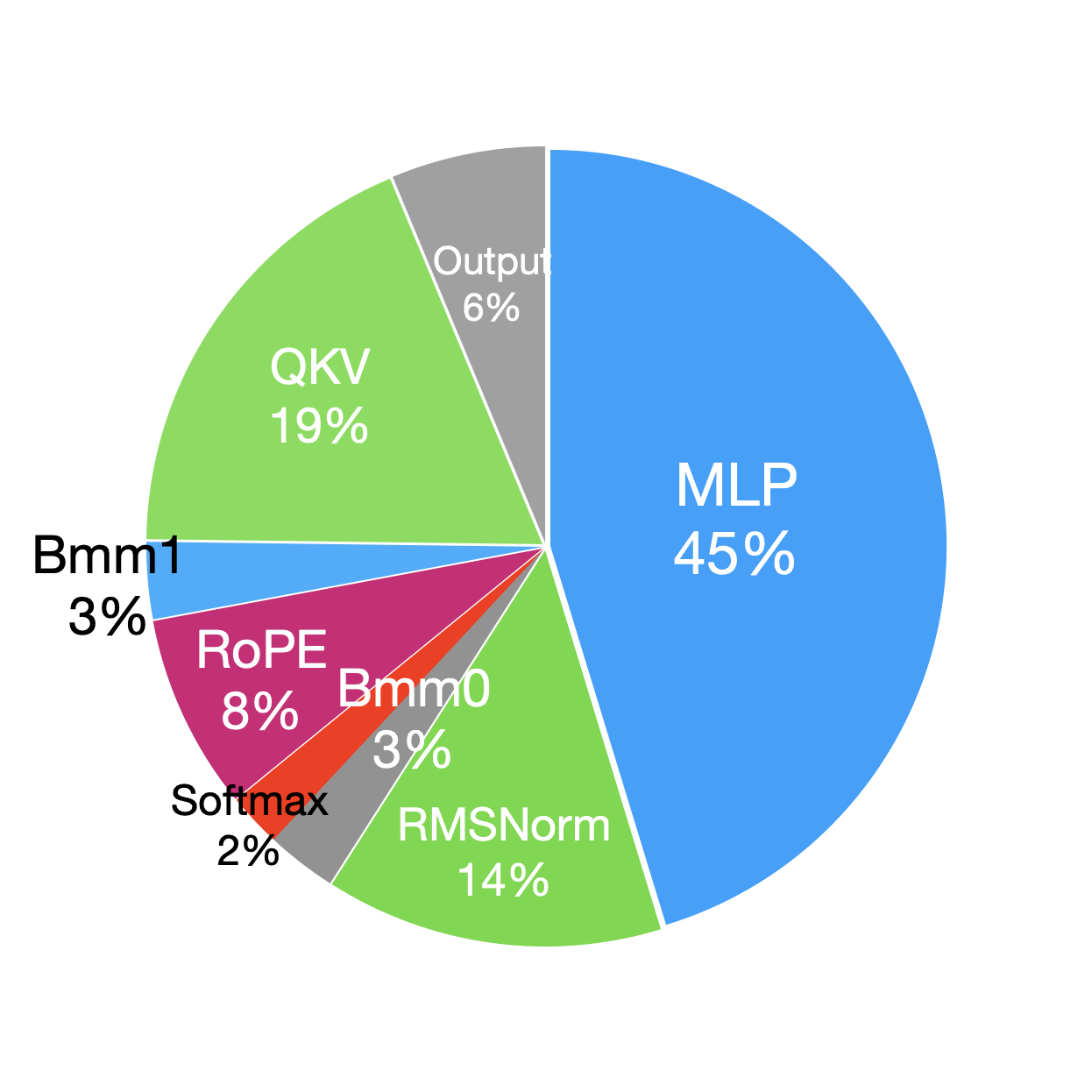}
        \captionsetup{justification=centering}
        \caption*{(c) Backward w/o recomputation}
     \end{subfigure}
    \begin{subfigure}[b]{0.18\textwidth}
        \centering
        \includegraphics[width=\linewidth]{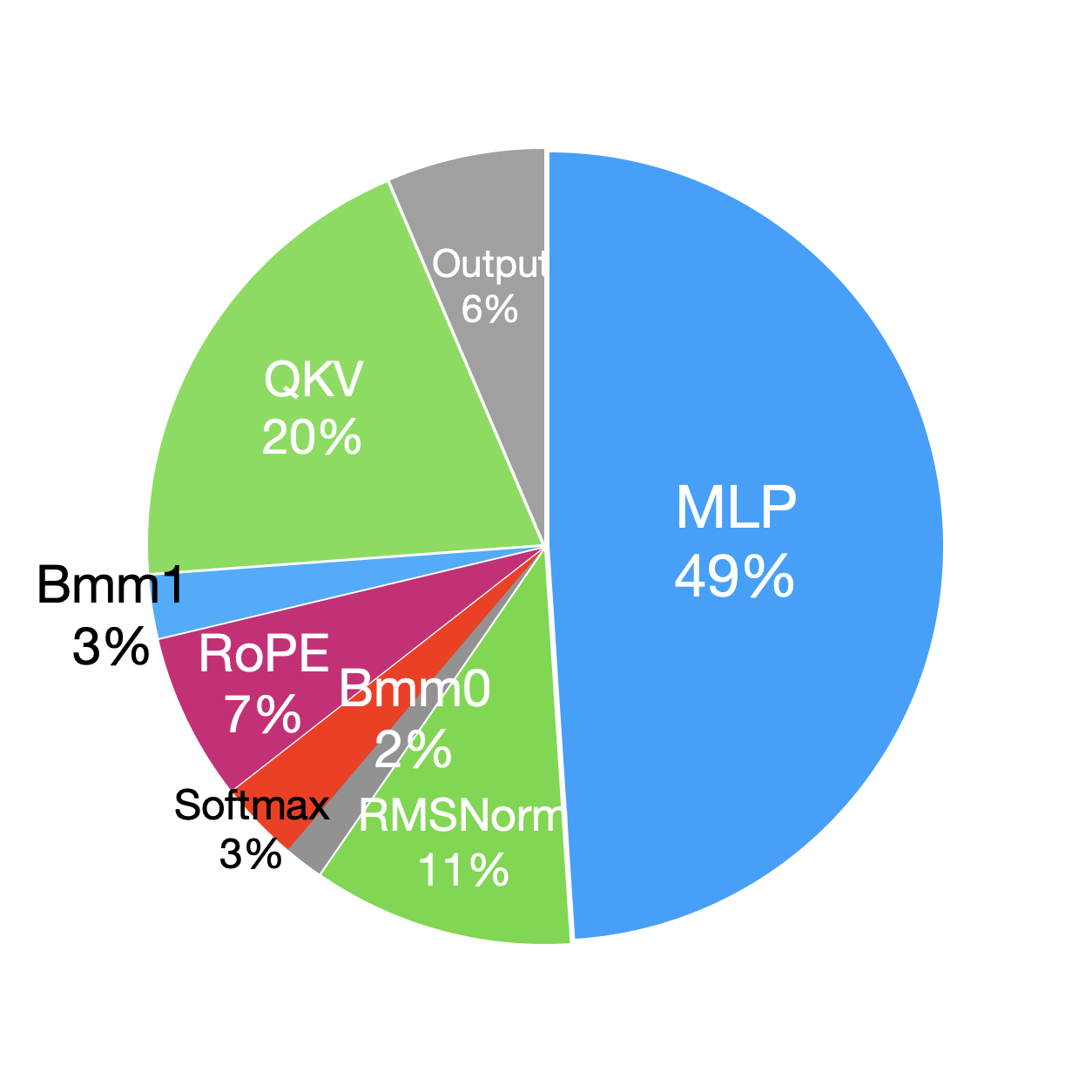}
        \captionsetup{justification=centering}
        \caption*{(d) Backward w/ recomputation}
     \end{subfigure}
\caption{Time breakdown of the decoder layer modules without (batch size 2) and with (batch size32) recomputation.}
     \label{fig:moduleproportion}
     \vspace{-10px}
\end{figure}

To further explore the impact of larger batch sizes on pre-training performance, we compare the percentage of time taken by decoder layer modules in the forward and backward (removing the recomputation part in backward) phases with and without using recomputation (i.e. comparison between batch size = 32 and batch size = 2). Because the recomputation part in the backward phase essentially reruns the forward phase, we analyze the forward and backward phases separately during profiling. Figure~\ref{fig:moduleproportion} shows that when the batch size increases from 2 to 32, the time breakdowns of modules in both the forward and backward phases do not change much. This is because, element-wise operations are memory-bound and their running time roughly scales linearly with batch size. In comparison, GEMM operations in the decoder layer are compute-bound, changing batch size often affects only one of M, N, or K, so the running time also grows linearly with batch size.

\begin{table}[!ht]
\centering
\caption{The overall and compute kernel time consumption of forward, backward and optimizer of one step pre-training Llama2-7B model.}
\begin{tabular}{c|ccc}
\hline
 & Forward & Backward\# & Optimizer \\ \hline
Overall (ms)                                                           & 900.8    & 2651.8    & 187.7   \\
Percentage(\%)                                                     & 24.0    & 70.8     & 5.1     \\
\begin{tabular}[c]{@{}c@{}}Compute kernel (ms)\end{tabular} & 895.0   & 2614.4   & 180.0  \\ \hline
\end{tabular}
\label{table:bs32proportion}
\vspace{-5px}
\end{table}

\textbf{FlashAttention} fuses the operations of $QK^T$, softmax, PV(P=softmax($\frac{QK^T}{\sqrt{d}}$)) and a few element-wise operations into one kernel, using more accesses to the low-latency high-bandwidth GPU SRAM and reducing accesses to the high-latency low-bandwidth GPU DRAM. Table \ref{table:attentionmodule} shows that this technique can accelerate attention module by 34.9\% and 24.7\% respectively.

\begin{table}[!ht]
\caption{Comparison of attention module time consumption in naive and FlashAttention methods.}
\centering
\begin{tabular}{c|c|c}
\hline
                   & Forward & backward \\ \hline
Naive(ms)          & 1.06    & 2.75     \\ \hline
FlashAttention(ms) & 0.69    & 2.07     \\ \hline
Improvement(\%)     & 34.9    & 24.7     \\ \hline
\end{tabular}
\label{table:attentionmodule}
\vspace{-15px}
\end{table}

\section{Results on Fine-tuning}\label{sec:exp-finetuning}

In fine-tuning, our primary focus is on parameter-efficient fine-tuning methods (PEFT methods), as full parameter training has already been discussed in Section~\ref{subsec:exp-endtoend}. We report on the fine-tuning performance of LoRA and QLoRA across various model sizes and hardware settings. We use a sequence length of 350, a batch size of 1, and a LoRA rank of 64, loading the model weight into bf16 by default. For QLoRA, we adopt a 4-bit configuration with double quantization\cite{dettmers2023qlora}. We also combine LoRA and QLoRA with other techniques, maintaining the same configurations as in Section~\ref{subsec:exp-endtoend}. The fine-tuning results for Llama2-7B are presented in Table~\ref{apx-table:finetuning-manymethod-bs1}.

Table \ref{apx-table:finetuning-manymethod-bs1} reveals that the performance trend in fine-tuning Llama2-13B using LoRA and QLoRA is consistent with that of Llama2-7B. Specifically, LoRA achieves approximately $2\times$ higher throughput than QLoRA in all evaluations, primarily due to the overhead associated with quantization and de-quantization operations. However, QLoRA's memory consumption is half that of LoRA's. FlashAttention and ZeRO-2, when combined with LoRA in fine-tuning, achieve 20\% and 10\% more throughput, respectively, than LoRA alone on all hardware platforms. In contrast, ZeRO-3 or offloading shows poor performance in LoRA fine-tuning, as LoRA updates only a small group of parameters, namely the low-rank adapters. Since the optimizer state is limited to handling LoRA parameter updates, which are not compute-bound, offloading or sharding such a small fraction of states introduces more communication overhead compared to computing time, and it does not significantly reduce memory usage.

The throughput for fine-tuning Llama2-13B shows a decrease of about 30\% compared to that of Llama2-7B. However, when all optimization techniques are combined, even RTX4090 and RTX3090 can fine-tune Llama2-70B, achieving a total throughput of around 200 tokens per second.

\begin{table*}[!ht]
\centering
\addtolength{\tabcolsep}{-4.5pt}
\caption{Fine-tuning performance comparision among LoRA (L), QLoRA (QL), with different optimization methods, including ZeRO stage 2 and 3 (Z2, Z3), FlashAttention (F), offloading (O), activation recomputation (R), on 4 types of 8-GPU servers: A800, RTX4090, RTX3090 w/ NVLink and RTX3090 w/o NVLink. In offloading, ZeRO-3 offloads both optimizer state and parameters to CPU, while ZeRO-2 only offloads optimizer state to CPU. We set the batch size to 1, and sequence length to 350. We report the average throughput tokens/s (Tokens/s) and its standard deviation in three independent runs and peak GPU memory usage (M) in GB. In each run, throughput is averaged over 100 steps after 30 warm-up steps.}
\begin{tabular}{l|l|cccccccc}
\hline
\multirow{3}{*}{Model} & \multirow{3}{*}{Method} & \multicolumn{8}{c}{Hardware platform} \\ \cline{3-10} 
 &  & \multicolumn{2}{c|}{A800} & \multicolumn{2}{c|}{RTX4090} & \multicolumn{2}{c|}{RTX3090 w/ NVLink} & \multicolumn{2}{c}{RTX3090 w/o NVLink} \\ \cline{3-10} 
 &  & Tokens/s & \multicolumn{1}{c|}{M (GB)} & Tokens/s & \multicolumn{1}{c|}{M (GB)} & Tokens/s & \multicolumn{1}{c|}{M (GB)} & Tokens/s & M (GB) \\ \hline
\multirow{18}{*}{7B} & L & $14216.6_{.03}$ & \multicolumn{1}{c|}{22.7} & $2875.3_{.03}$ & \multicolumn{1}{c|}{20.5} & $1936._{.03}$ & \multicolumn{1}{c|}{20.5} & $1866.3_{.02}$ & 20.5 \\
 & QL & $7631.2_{.02}$ & \multicolumn{1}{c|}{13.7} & $2151._{.02}$ & \multicolumn{1}{c|}{14} & $1602._{.01}$ & \multicolumn{1}{c|}{14} & $1359.8_{.01}$ & 14 \\
 & L+R & $11202.7_{.05}$ & \multicolumn{1}{c|}{21.9} & $2410.1_{.01}$ & \multicolumn{1}{c|}{20.1} & $1636.4_{.02}$ & \multicolumn{1}{c|}{20.1} & $1609._{.03}$ & 20.1 \\
 & QL+R & $5186.4_{.07}$ & \multicolumn{1}{c|}{11} & $1947.6_{.03}$ & \multicolumn{1}{c|}{11.9} & $1397.3_{.03}$ & \multicolumn{1}{c|}{11.9} & $1384.5_{.0}$ & 11.9 \\
 & L+F & $17182._{.1}$ & \multicolumn{1}{c|}{20.5} & $3245.2_{.01}$ & \multicolumn{1}{c|}{18.9} & $2278.8_{.03}$ & \multicolumn{1}{c|}{18.9} & $2272.7_{.02}$ & 18.9 \\
 & QL+F & $9792.5_{.04}$ & \multicolumn{1}{c|}{9.5} & $3378.3_{.02}$ & \multicolumn{1}{c|}{10.5} & $2524.4_{.02}$ & \multicolumn{1}{c|}{10.5} & $2514.4_{.02}$ & 10.5 \\
 & L+Z2 & $15734.1_{.04}$ & \multicolumn{1}{c|}{19.1} & $4118.6_{.03}$ & \multicolumn{1}{c|}{19} & $3207._{.02}$ & \multicolumn{1}{c|}{19} & $3034.4_{.02}$ & 19 \\
 & L+Z2+O & $9152.4_{.02}$ & \multicolumn{1}{c|}{18.8} & $2761.9_{.01}$ & \multicolumn{1}{c|}{18.7} & $2168.3_{.01}$ & \multicolumn{1}{c|}{18.7} & $1909.9_{.01}$ & 18.7 \\
 & L+Z3 & $2846.1_{.01}$ & \multicolumn{1}{c|}{13.3} & $225.3_{.0}$ & \multicolumn{1}{c|}{13.3} & $160.9_{.0}$ & \multicolumn{1}{c|}{13.3} & $155.7_{.0}$ & 13.3 \\
 & L+Z3+O & $1878.3_{.0}$ & \multicolumn{1}{c|}{11.2} & $195.2_{.0}$ & \multicolumn{1}{c|}{11.4} & $131.8_{.0}$ & \multicolumn{1}{c|}{11.4} & $129.1_{.0}$ & 11.4 \\
 & QL+Z2 & $10074.3_{.04}$ & \multicolumn{1}{c|}{10.6} & $2105.7_{.02}$ & \multicolumn{1}{c|}{10.5} & $1471.1_{.02}$ & \multicolumn{1}{c|}{10.5} & $1443.6_{.03}$ & 10.5 \\
 & QL+Z2+O & $6700.1_{.03}$ & \multicolumn{1}{c|}{10.3} & $1814.3_{.01}$ & \multicolumn{1}{c|}{10.3} & $1417._{.01}$ & \multicolumn{1}{c|}{10.3} & $1274.7_{.03}$ & 10.3 \\
 & L+F+R & $12906.3_{.05}$ & \multicolumn{1}{c|}{22.2} & $3779.5_{.02}$ & \multicolumn{1}{c|}{18.9} & $2777.5_{.02}$ & \multicolumn{1}{c|}{18.9} & $2769.7_{.02}$ & 18.9 \\
 & QL+F+R & $6864.3_{.05}$ & \multicolumn{1}{c|}{8.5} & $2088.4_{.03}$ & \multicolumn{1}{c|}{10.1} & $1528.4_{.04}$ & \multicolumn{1}{c|}{10.1} & $1506._{.03}$ & 10.1 \\
 & L+F+R+Z2 & $12730.3_{.12}$ & \multicolumn{1}{c|}{15.6} & $3222.8_{.03}$ & \multicolumn{1}{c|}{15.5} & $2258.2_{.03}$ & \multicolumn{1}{c|}{15.5} & $2194.7_{.03}$ & 15.5 \\
 & L+F+R+Z2+O & $8001.8_{.03}$ & \multicolumn{1}{c|}{15.3} & $2525.3_{.02}$ & \multicolumn{1}{c|}{15.2} & $1778.6_{.04}$ & \multicolumn{1}{c|}{15.2} & $1670.1_{.02}$ & 15.2 \\
 & L+F+R+Z3 & $2395.7_{.01}$ & \multicolumn{1}{c|}{8.5} & $222.1_{.0}$ & \multicolumn{1}{c|}{9.3} & $162.2_{.0}$ & \multicolumn{1}{c|}{9.3} & $156.6_{.0}$ & 9.3 \\
 & L+F+R+Z3+O & $1691.1_{.05}$ & \multicolumn{1}{c|}{7} & $199.5_{.0}$ & \multicolumn{1}{c|}{7.7} & $143.1_{.0}$ & \multicolumn{1}{c|}{7.7} & $166.5_{.0}$ & 7.7 \\ \hline
\multirow{18}{*}{13B} & L & $9724.5_{.17}$ & \multicolumn{1}{c|}{40.3} & \multicolumn{2}{c|}{-} & \multicolumn{2}{c|}{-} & \multicolumn{2}{c}{-} \\
 & QL & $4524.5_{.23}$ & \multicolumn{1}{c|}{21.7} & $1449.5_{.03}$ & \multicolumn{1}{c|}{21.7} & $1127.9_{.03}$ & \multicolumn{1}{c|}{21.7} & $868.6_{.03}$ & 21.7 \\
 & L+R & $7236.8_{.05}$ & \multicolumn{1}{c|}{36.5} & \multicolumn{2}{c|}{-} & \multicolumn{2}{c|}{-} & \multicolumn{2}{c}{-} \\
 & QL+R & $3044.1_{.03}$ & \multicolumn{1}{c|}{15.5} & $1234.2_{.02}$ & \multicolumn{1}{c|}{18.5} & $933.4_{.04}$ & \multicolumn{1}{c|}{18.5} & $738.2_{.04}$ & 18.5 \\
 & L+F & $11867.2_{.1}$ & \multicolumn{1}{c|}{41.4} & \multicolumn{2}{c|}{-} & \multicolumn{2}{c|}{-} & \multicolumn{2}{c}{-} \\
 & QL+F & $5657.8_{.03}$ & \multicolumn{1}{c|}{15.7} & $2343.5_{.02}$ & \multicolumn{1}{c|}{16.8} & $1577.7_{.05}$ & \multicolumn{1}{c|}{16.8} & $1377._{.01}$ & 16.8 \\
 & L+Z2 & $11718.2_{.1}$ & \multicolumn{1}{c|}{33.8} & \multicolumn{2}{c|}{-} & \multicolumn{2}{c|}{-} & \multicolumn{2}{c}{-} \\
 & L+Z2+O & $5997.1_{.04}$ & \multicolumn{1}{c|}{33.5} & \multicolumn{2}{c|}{-} & \multicolumn{2}{c|}{-} & \multicolumn{2}{c}{-} \\
 & L+Z3 & $2271.3_{.11}$ & \multicolumn{1}{c|}{18.1} & $109.9_{.0}$ & \multicolumn{1}{c|}{17.9} & $92.8_{.0}$ & \multicolumn{1}{c|}{17.9} & $77.5_{.0}$ & 17.9 \\
 & L+Z3+O & $1351.6_{.07}$ & \multicolumn{1}{c|}{14.1} & $102.1_{.0}$ & \multicolumn{1}{c|}{14.2} & $88._{.0}$ & \multicolumn{1}{c|}{14.2} & $70.5_{.0}$ & 14.2 \\
 & QL+Z2 & $5958.8_{.03}$ & \multicolumn{1}{c|}{16.7} & $2171._{.03}$ & \multicolumn{1}{c|}{16.8} & $1541.4_{.05}$ & \multicolumn{1}{c|}{16.8} & $1322.3_{.05}$ & 16.8 \\
 & QL+Z2+O & $4064.7_{.02}$ & \multicolumn{1}{c|}{16.3} & $1734.1_{.02}$ & \multicolumn{1}{c|}{16.4} & $1402.3_{.02}$ & \multicolumn{1}{c|}{16.4} & $1028.8_{.02}$ & 16.4 \\
 & L+F+R & $9005.3_{.1}$ & \multicolumn{1}{c|}{40.3} & \multicolumn{2}{c|}{-} & \multicolumn{2}{c|}{-} & \multicolumn{2}{c}{-} \\
 & QL+F+R & $4136.8_{.06}$ & \multicolumn{1}{c|}{13.9} & $1512.9_{.3}$ & \multicolumn{1}{c|}{16.8} & $1146._{.02}$ & \multicolumn{1}{c|}{16.8} & $905._{.02}$ & 16.8 \\
 & L+F+R+Z2 & $9826.9_{.16}$ & \multicolumn{1}{c|}{28.1} & \multicolumn{2}{c|}{-} & \multicolumn{2}{c|}{-} & \multicolumn{2}{c}{-} \\
 & L+F+R+Z2+O & $5534.8_{.14}$ & \multicolumn{1}{c|}{27.8} & \multicolumn{2}{c|}{-} & \multicolumn{2}{c|}{-} & \multicolumn{2}{c}{-} \\
 & L+F+R+Z3 & $1912.9_{.0}$ & \multicolumn{1}{c|}{10.8} & $114.2_{.0}$ & \multicolumn{1}{c|}{11.9} & $97.3_{.0}$ & \multicolumn{1}{c|}{11.9} & $81.9_{.0}$ & 11.9 \\
 & L+F+R+Z3+O & $1226.7_{.02}$ & \multicolumn{1}{c|}{8.7} & $101.1_{.0}$ & \multicolumn{1}{c|}{8.8} & $91._{.0}$ & \multicolumn{1}{c|}{8.8} & $78.5_{.0}$ & 8.8 \\ \hline
\multirow{5}{*}{70B} & QL+F+R & $1043.8_{.01}$ & \multicolumn{1}{c|}{59.8} & \multicolumn{2}{c|}{-} & \multicolumn{2}{c|}{-} & \multicolumn{2}{c}{-} \\
 & L+F+R+Z3 & $967.3_{.02}$ & \multicolumn{1}{c|}{29.4} & \multicolumn{2}{c|}{-} & \multicolumn{2}{c|}{-} & \multicolumn{2}{c}{-} \\
 & L+F+R+Z3+O & $427._{.0}$ & \multicolumn{1}{c|}{13.2} & $19.4_{.0}$ & \multicolumn{1}{c|}{13.2} & $12._{.0}$ & \multicolumn{1}{c|}{13.2} & $10.1_{.0}$ & 13.2 \\
 & QL+R & $717.1_{.0}$ & \multicolumn{1}{c|}{63.6} & \multicolumn{2}{c|}{-} & \multicolumn{2}{c|}{-} & \multicolumn{2}{c}{-} \\
 & QL+F & $1385.7_{.03}$ & \multicolumn{1}{c|}{61.2} & \multicolumn{2}{c|}{-} & \multicolumn{2}{c|}{-} & \multicolumn{2}{c}{-} \\ \hline
\end{tabular}
\label{apx-table:finetuning-manymethod-bs1}
\end{table*}

\section{Results on Inference}\label{sec:exp-inference}
\subsection{End-to-end Performance}\label{subsec:inference-endtoend}

\begin{figure}[!ht]
     \centering
     \begin{subfigure}[b]{0.155\textwidth}
        \centering
        \includegraphics[width=\linewidth]{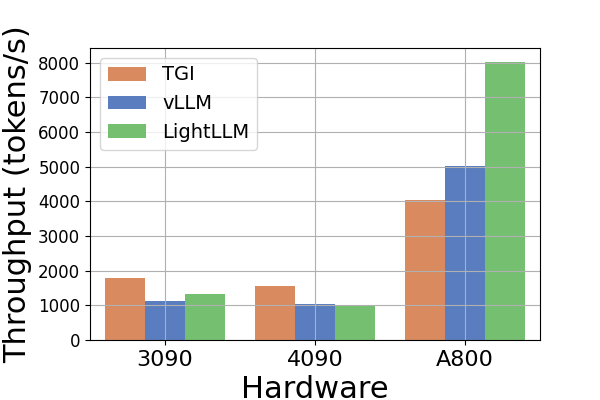}
        \caption*{(a) Llama2-7B}
        \label{fig:different_machine_compare_7b}
     \end{subfigure}
     \begin{subfigure}[b]{0.155\textwidth}
         \centering
        \includegraphics[width=\linewidth]{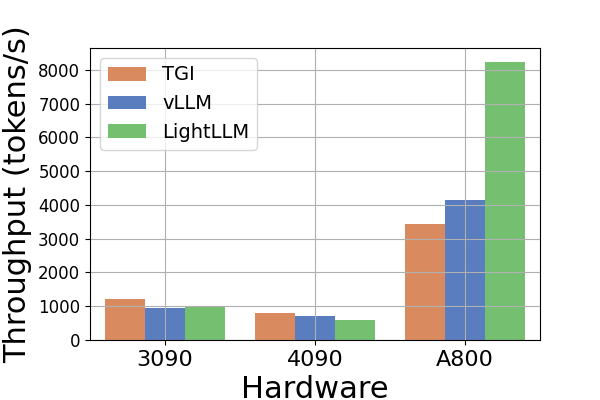}
        \caption*{(b) Llama2-13B}
        \label{fig:different_machine_compare_13b}
     \end{subfigure}
     \begin{subfigure}[b]{0.155\textwidth}
         \centering
        \includegraphics[width=\linewidth]{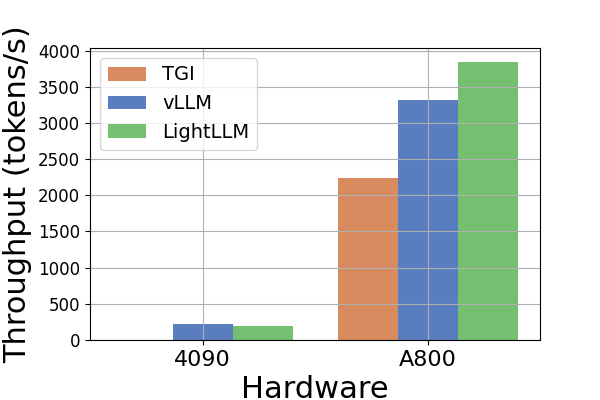}
        \caption*{(c) Llama2-70B}
        \label{fig:different_machine_compare_70b}
     \end{subfigure}
        \caption{Throughput performance comparison on different GPU platforms with LLM size varied.}
        \label{fig:different_machine_compare}
\end{figure}

\textit{Throughput.} A comparative analysis of throughput across various hardware platforms and inference frameworks is presented in Figure~\ref{fig:different_machine_compare}. As the Llama2-70B model induces an OOM error with the TGI framework on RTX3090 and RTX4090, related data inference to Llama2-70B is omitted in Figure~\ref{fig:different_machine_compare}. The TGI framework demonstrates superior throughput, particularly on GPUs with 24GB memory, such as RTX3090 and RTX4090. In comparison, LightLLM significantly outperforms TGI and vLLM on the A800 GPU platform by nearly doubling the throughput. These experiments reveal that the TGI inference framework yields superior performance on 24GB GPU platforms, whereas the LightLLM inference framework exhibits the highest throughput on the A800 80GB GPU platform. This finding shows that LightLLM is specifically optimized for high-performance GPUs such as the A800/A100 series.

\begin{figure}[!ht]
    \vspace{-8px}
     \centering
     \begin{subfigure}[b]{0.23\textwidth}
        \centering
        \includegraphics[width=\linewidth]{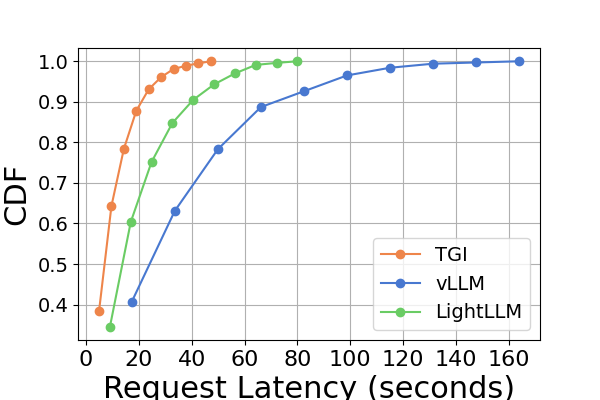}
        \phantomcaption
        \caption*{(a) Llama2-7B with RTX3090}
        \label{fig:latency_same_machine_compare_3090_7b}
     \end{subfigure}
     \begin{subfigure}[b]{0.23\textwidth}
         \centering
        \includegraphics[width=\linewidth]{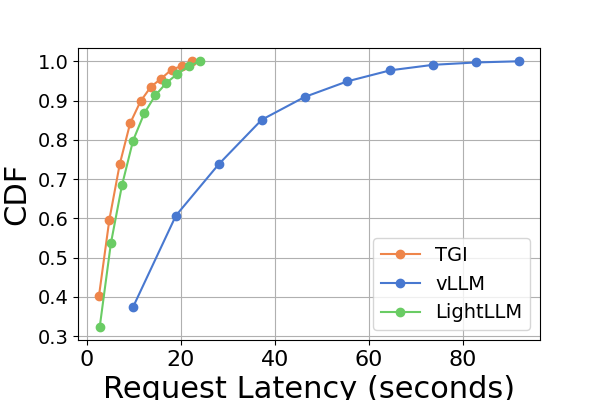}
        \phantomcaption
        \caption*{(b) Llama2-7B with A800}
        \label{fig:latency_same_machine_compare_A800_7b}
     \end{subfigure}
     \caption{Latency comparison on the same GPU platform with different inference frameworks. }
        \label{fig:latency_same_machine_compare}
\end{figure}

\begin{figure}[!ht]
     \centering
     \begin{subfigure}[b]{0.155\textwidth}
         \centering
        \includegraphics[width=\linewidth]{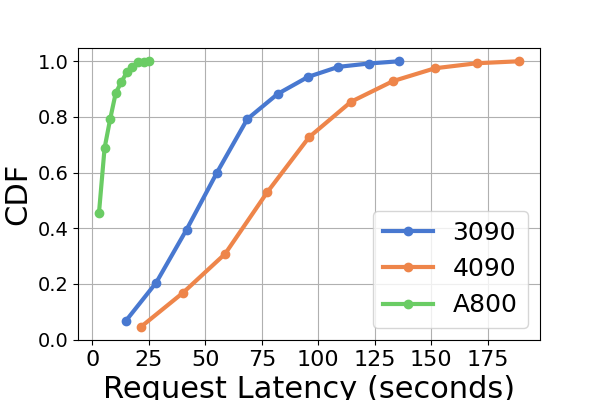}
        \caption*{%
            \begin{tabular}{c}
            (a) TGI
        \end{tabular}}        
        \label{fig:latency_same_framework_compare_TGI_13b}
     \end{subfigure}
     \begin{subfigure}[b]{0.155\textwidth}
         \centering
        \includegraphics[width=\linewidth]{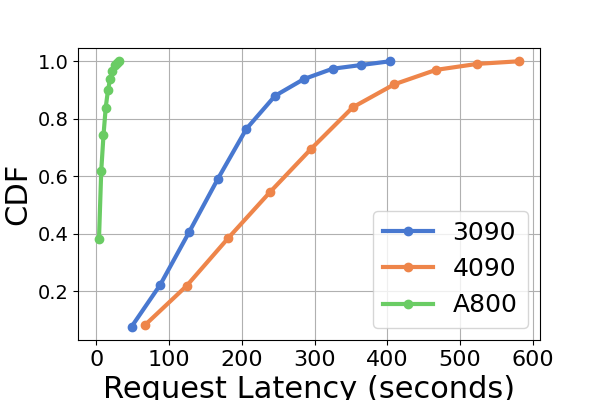}
        \caption*{%
            \begin{tabular}{c}
            (b) vLLM
        \end{tabular}}        
        \label{fig:latency_same_framework_compare_vLLM_13b}
     \end{subfigure}
     \begin{subfigure}[b]{0.155\textwidth}
         \centering
        \includegraphics[width=\linewidth]{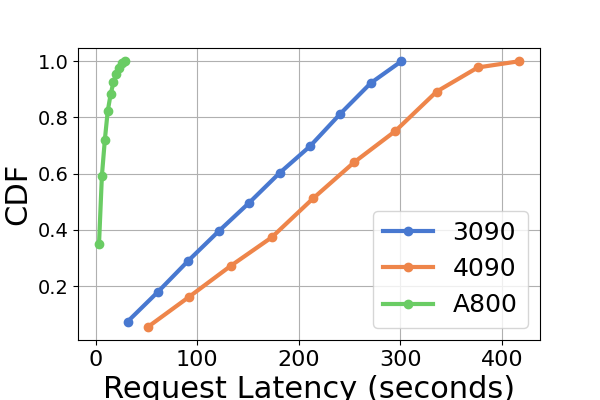}
        \caption*{%
            \begin{tabular}{c}
            (c) LightLLM
        \end{tabular}}        
        \label{fig:latency_same_framework_compare_LightLLM_13b}
     \end{subfigure}
     \caption{Latency performance comparison on different GPU platforms with Llama2-13B.}
    \label{fig:latency_same_framework_compare}
\end{figure}

\begin{figure}[!ht]
     \centering
      \begin{subfigure}[b]{0.23\textwidth}
         \centering
        \includegraphics[width=\linewidth]{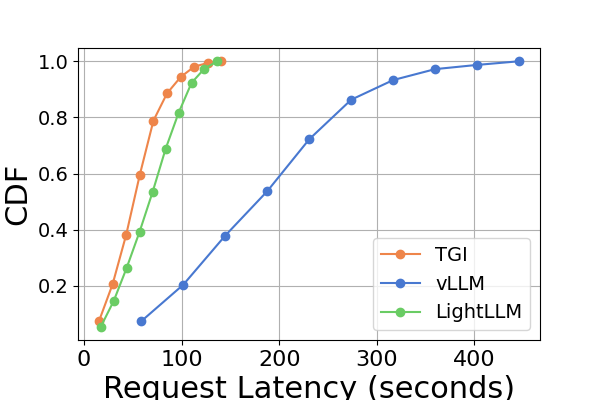}
        \caption*{(a) Llama2-13B with 3090}
        \label{fig:latency_same_machine_compare_3090_13b}
     \end{subfigure}
     \begin{subfigure}[b]{0.23\textwidth}
         \centering
        \includegraphics[width=\linewidth]{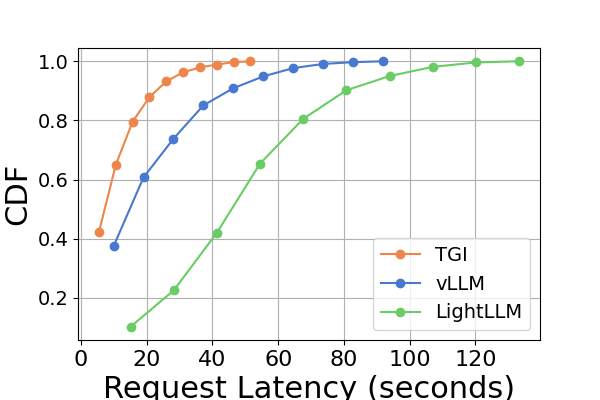}
        \caption*{(b) Llama2-7B with 4090}
        \label{fig:latency_same_machine_compare_4090_7b}
     \end{subfigure}
     \begin{subfigure}[b]{0.23\textwidth}
         \centering
        \includegraphics[width=\linewidth]{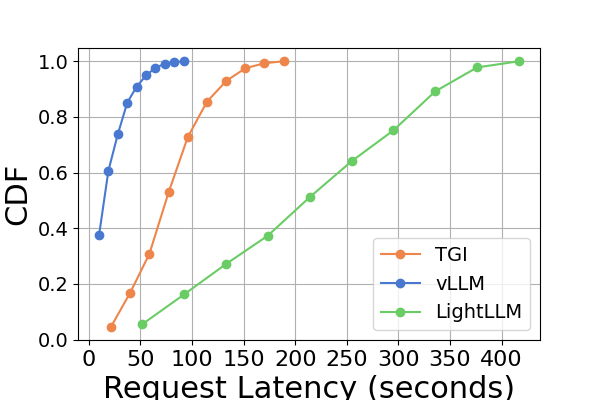}
        \caption*{(c) Llama2-13B with 4090}
        \label{fig:latency_same_machine_compare_4090_13b}
     \end{subfigure}
     \begin{subfigure}[b]{0.23\textwidth}
         \centering
        \includegraphics[width=\linewidth]{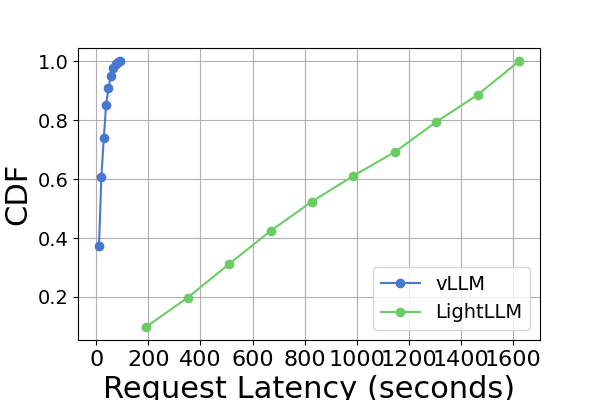}
        \caption*{(d) Llama2-70B with 4090}
        \label{fig:latency_same_machine_compare_4090_70b}
     \end{subfigure}
     \begin{subfigure}[b]{0.23\textwidth}
         \centering
        \includegraphics[width=\linewidth]{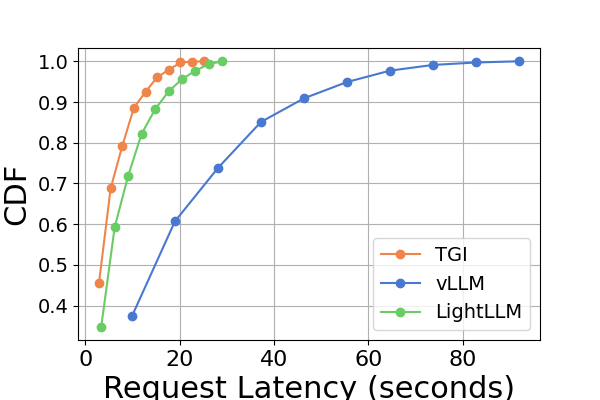}
        \caption*{(e) Llama2-13B with A800}
        \label{fig:latency_same_machine_compare_A800_13b}
     \end{subfigure}
      \begin{subfigure}[b]{0.23\textwidth}
         \centering
        \includegraphics[width=\linewidth]{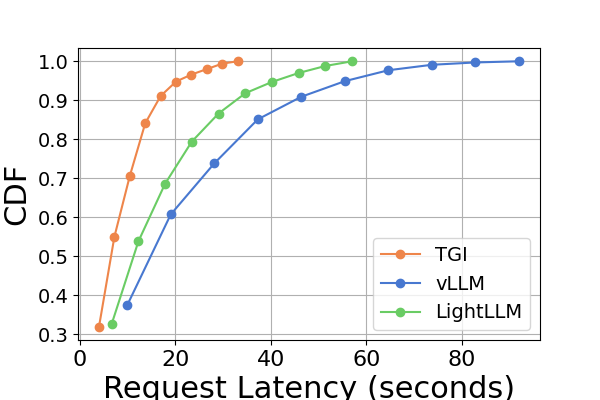}
        \caption*{(f) Llama2-70B with A800}
        \label{fig:latency_same_machine_compare_A800_70b}
     \end{subfigure}
     \caption{Latency performance comparison on the same GPU platform with different inference frameworks.}
        \label{fig:latency_same_machine_compare_appendix}
\end{figure}

\begin{figure}[!ht]
     \centering
     \begin{subfigure}[b]{0.23\textwidth}
        \centering
        \includegraphics[width=\linewidth]{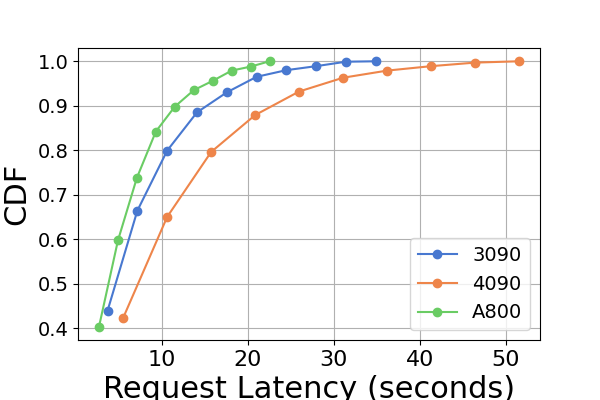}
        \caption*{%
            \begin{tabular}{c}
            (a) Llama2-7B with\\
            TGI
        \end{tabular}}        
        \label{fig:latency_same_framework_compare_TGI_7b}
     \end{subfigure}
     \begin{subfigure}[b]{0.23\textwidth}
         \centering
        \includegraphics[width=\linewidth]{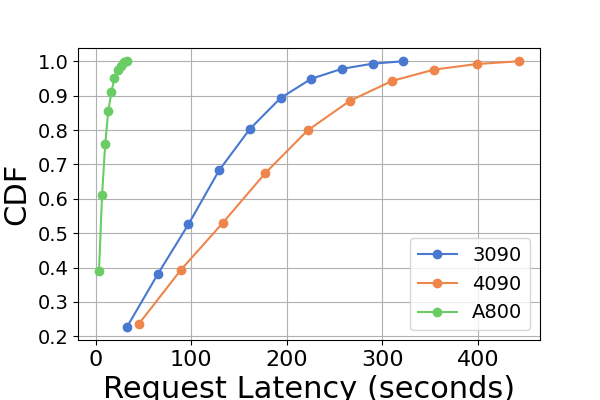}
        \caption*{%
            \begin{tabular}{c}
            (b) Llama2-7B with\\
            vLLM
        \end{tabular}}        
        \label{fig:latency_same_framework_compare_vLLM_7b}
     \end{subfigure}
     \begin{subfigure}[b]{0.23\textwidth}
         \centering
        \includegraphics[width=\linewidth]{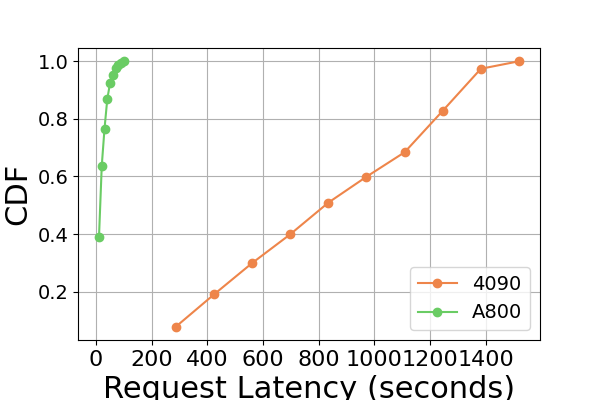}
        \caption*{%
            \begin{tabular}{c}
            (c) Llama2-70B with\\
            vLLM
        \end{tabular}}        
        \label{fig:latency_same_framework_compare_vLLM_70b}
     \end{subfigure}
     \begin{subfigure}[b]{0.23\textwidth}
         \centering
        \includegraphics[width=\linewidth]{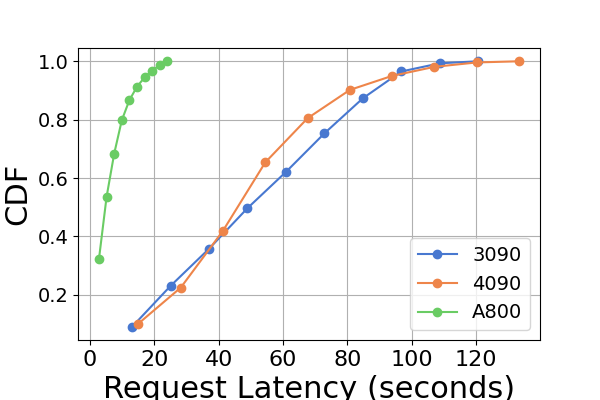}
        \caption*{%
            \begin{tabular}{c}
            (d) Llama2-7B with\\
            LightLLM
        \end{tabular}}
        \label{fig:latency_same_framework_compare_LightLLM_7b}
     \end{subfigure}
     \begin{subfigure}[b]{0.23\textwidth}
         \centering
        \includegraphics[width=\linewidth]{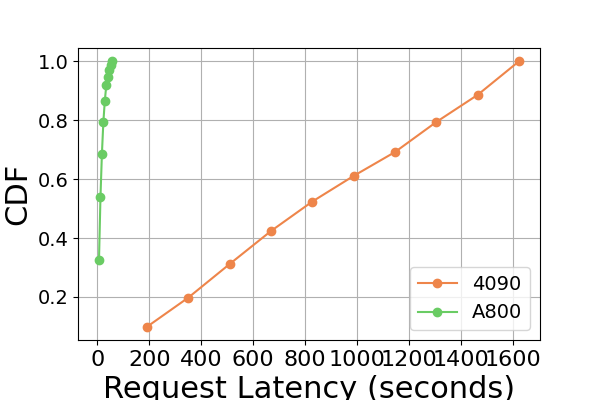}
        \caption*{%
            \begin{tabular}{c}
            (e) Llama2-70B with\\
            LightLLM
        \end{tabular}}        
        \label{fig:latency_same_framework_compare_LightLLM_70b}
     \end{subfigure}
     \caption{Latency performance comparison on different GPU platforms with the same inference framework. CDF stands for Cumulative Distribution Function.}
        \label{fig:latency_same_framework_compare_appendix}
\end{figure}

\textit{Latency.} A comparative analysis of latency across various hardware platforms and inference frameworks is presented in Figure~\ref{fig:latency_same_machine_compare} and Figure~\ref{fig:latency_same_framework_compare} and Figure~\ref{fig:latency_same_machine_compare_appendix} and Figure~\ref{fig:latency_same_framework_compare_appendix}. We use the Cumulative Distribution Function (CDF) to plot the latency across different inference frameworks. The CDF represents the probability that a variable takes on a value less than or equal to a particular point in the sample space. For example, in Figure~\ref{fig:latency_same_machine_compare_3090_7b} illustrates that LightLLM takes approximately 20 seconds to respond to 60\% of the requests, and requires around 80 seconds to respond to 100\% of the requests.

In Figure~\ref{fig:latency_same_machine_compare}, we compare the latency of three inference frameworks specifically on the same GPU platform. 
The performance on RTX3090 and A800 platforms exhibits similar trends, with TGI showing the lowest latency, followed by LightLLM, and vLLM having the highest latency. Additional experiments are presented in Figure~\ref{fig:latency_same_machine_compare_appendix}.

Figure~\ref{fig:latency_same_machine_compare_appendix} is an extension to Figure~\ref{fig:latency_same_machine_compare}, containing additional latency experiments on the same GPU platform for the inference benchmark. The performance results on the RTX4090 platform are different from the other two platforms. This discrepancy might be due to the \texttt{NCCL\_P2P\_DISABLE=1} setting. On the RTX4090 platform, LightLLM shows the highest latency, and TGI has the lowest latency for the Llama2-7B model.
Another distinct finding from the latency experiment is that on the consumer-level GPU platforms the total inference time increases as the model's parameter size grows. Specifically on the RTX4090 platform, the inference time difference between Llama2-7B and Llama2-70B can reach up to 13 times, from 120 seconds to 1600 seconds. However, this phenomenon is not observed on the A800 GPU platform, where the inference time for larger models remains within a narrow range. This indicates that for currently popular LLM sizes, the A800 platform can handle inference without any latency implications, and that the 70B model has not reached the performance limit of the A800 platform for inference.

In Figure~\ref{fig:latency_same_framework_compare} we compare the latency of each inference framework across different GPU platforms. A800 consistently exhibits the lowest latency in almost all experiments. Furthermore, the RTX3090 GPU platform demonstrates a lower latency than the RTX4090 in the majority of experiments, a situation that might also result from the \texttt{NCCL\_P2P\_DISABLE=1} setting. These experiments suggest that if one aims for an inference service with the least latency, the A800 GPU platform is the best choice, offering significant performance advantages across various model and inference framework combinations. Additional experiments are presented in Figure~\ref{fig:latency_same_framework_compare_appendix}.

Figure~\ref{fig:latency_same_framework_compare_appendix} is an extension to Figure~\ref{fig:latency_same_framework_compare}, containing additional latency experiments on the same inference framework for the inference benchmark. 

In summary, the A800 platform significantly outperforms the other two consumer-level platforms in both throughput and latency. Among the two consumer-level platforms, the RTX3090 has a slight advantage over the RTX4090. When operating on a consumer-level platform, the three inference frameworks do not show a substantial difference in terms of throughput. In contrast, the TGI framework consistently outperforms the others in latency. On the A800 GPU platform, LightLLM is the top performer on throughput, and its latency is also remarkably close to the TGI framework.

\vspace*{-0.5cm}

\subsection{Module-wise Analysis}\label{subsec:inference-modulewise}
In this subsection, we discuss the module-wise time cost using LightLLM as an example. To simulate a scenario with a large number of users accessing the service, we set the batch size to 1024, the output length to 64, and the prompt length to 512 on the A800 GPU server. The results are presented in Table~\ref{tab:module_microbenchmark_1} and Table~\ref{tab:module_microbenchmark_2}. We observe that the GPU experiences a bottleneck, evident in the 'Other' row of Table~\ref{tab:module_microbenchmark_1}, which accounts for 7.55\% of the total time. This suggests that fusing GPU kernels could potentially reduce the overall duration of this bottleneck.

\begin{table}[htbp]
\centering
\caption{The module-wise time cost of Llama2-7B running LightLLM on A800. }
    \begin{tabular}{c|cc|cc|cc}
    \hline
                                       & \multicolumn{2}{c|}{Task}                        & \multicolumn{2}{c|}{Time(ms)}             & \multicolumn{2}{c}{Percentage(\%)}         \\ \hline
    \multirow{10}{*}{Forward} & \multicolumn{1}{c|}{\multirow{6}{*}{Comp.}} & Element-Wise & \multicolumn{1}{c|}{\multirow{6}{*}{36}}   & 1.69  & \multicolumn{1}{c|}{\multirow{6}{*}{70.44}} & 3.3   \\ \cline{3-3} \cline{5-5} \cline{7-7} 
                                       & \multicolumn{1}{c|}{}                       & RoPE       & \multicolumn{1}{c|}{}                      & 0.19  & \multicolumn{1}{c|}{}                       & 0.37  \\ \cline{3-3} \cline{5-5} \cline{7-7} 
                                       & \multicolumn{1}{c|}{}                       & Triton       & \multicolumn{1}{c|}{}                      & 23.05 & \multicolumn{1}{c|}{}                       & 45.1  \\ \cline{3-3} \cline{5-5} \cline{7-7} 
                                       & \multicolumn{1}{c|}{}                       & GeMM         & \multicolumn{1}{c|}{}                      & 9.4   & \multicolumn{1}{c|}{}                       & 18.4  \\ \cline{3-3} \cline{5-5} \cline{7-7} 
                                       & \multicolumn{1}{c|}{}                       & RMSNorm      & \multicolumn{1}{c|}{}                      & 1.18  & \multicolumn{1}{c|}{}                       & 2.31  \\ \cline{3-3} \cline{5-5} \cline{7-7} 
                                       & \multicolumn{1}{c|}{}                       & Other        & \multicolumn{1}{c|}{}                      & 0.49  & \multicolumn{1}{c|}{}                       & 0.96  \\ \cline{2-7} 
                                       & \multicolumn{1}{c|}{\multirow{3}{*}{Comm.}} & AllReduce    & \multicolumn{1}{c|}{\multirow{3}{*}{11.3}} & 10.74 & \multicolumn{1}{c|}{\multirow{3}{*}{22.11}} & 21.01 \\ \cline{3-3} \cline{5-5} \cline{7-7} 
                                       & \multicolumn{1}{c|}{}                       & AllGather    & \multicolumn{1}{c|}{}                      & 0.46  & \multicolumn{1}{c|}{}                       & 0.9   \\ \cline{3-3} \cline{5-5} \cline{7-7} 
                                       & \multicolumn{1}{c|}{}                       & Other        & \multicolumn{1}{c|}{}                      & 0.1   & \multicolumn{1}{c|}{}                       & 0.2   \\ \cline{2-7} 
                                       & \multicolumn{2}{c|}{Other}                                 & \multicolumn{2}{c|}{3.81}                          & \multicolumn{2}{c}{7.55}                             \\ \hline
    \end{tabular}
\label{tab:module_microbenchmark_1}
\end{table}

\begin{table}[htbp]
\centering
\caption{The timeline of Llama2-7B using LightLLM on A800.}
    \begin{tabular}{cc|cc|cl}
    \hline
    \multicolumn{2}{c|}{Timeline}                             & \multicolumn{2}{c|}{Time(ms)}              & \multicolumn{2}{c}{Percentage(\%)}         \\ \hline
    \multicolumn{2}{c|}{Before Transformer}                          & \multicolumn{2}{c|}{1.66}                           & \multicolumn{2}{c}{3.25}                            \\ \hline
    \multicolumn{1}{c|}{\multirow{2}{*}{Transformer}} & 32 x Attention & \multicolumn{1}{c|}{\multirow{2}{*}{47.60}} & 35.13 & \multicolumn{1}{c|}{\multirow{2}{*}{93.13}} & 68.73 \\ \cline{2-2} \cline{4-4} \cline{6-6} 
    \multicolumn{1}{c|}{}                             & 32 x FFN       & \multicolumn{1}{c|}{}                       & 12.47 & \multicolumn{1}{c|}{}                       & 24.4  \\ \hline
    \multicolumn{2}{c|}{After Transformer}                           & \multicolumn{2}{c|}{1.85}                           & \multicolumn{2}{c}{3.62}                            \\ \hline
    \end{tabular}
\label{tab:module_microbenchmark_2}
\vspace{-10px}
\end{table}

\section{Microbenchmarks}\label{subsec:exp-microbenchmarks}
In order to get a deeper understanding on the experimental results, we conducted a microbenchmark analysis covering both computations and communications.

\subsection{GEMM Analysis} 

\begin{figure}[!ht]
\centering
\includegraphics[width=0.8\linewidth]{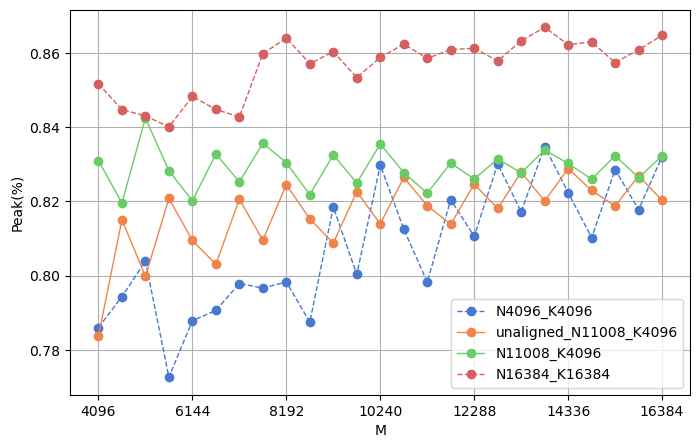}
\caption{GEMM performance with matrix sizes varied.}
\label{fig:GEMMpeak}
\vspace{-20px}
\end{figure}

In Section \ref{subsec:exp-modulewise}, we observe that the time consumption for modules containing GEMM operations is relatively high. We calculate the time breakdown of GEMM operations in the forward and backward phases using both the naive method and recomputation. Table \ref{table:Gemmproportion} shows that the GEMM kernel accounts for over 60\% of the time in both phases, highlighting the critical nature of GEMM performance for LLMs. To better understand GEMM performance, we analyze the first GEMM in the MLP module (Table \ref{table:GEMMPerf}). We choose this GEMM operation because MLP is the most time-consuming module, containing three GEMMs of similar sizes. The main reason for the lower peak values with the naive method compared to recomputation is the small matrix size, which fails to fully utilize the hardware. After increasing the size by 32 times using recomputation, the peak performance is still lower than the ideal value of 90\%. We test different matrix sizes on our experimental platform (A800), and the results are shown in Figure \ref{fig:GEMMpeak}. In these GEMM operations, the batch size affects $M$. Since we gradually increase $M$ under constant $K$ and $N$ values, there are two scenarios for choosing $M$. For N4096\_K4096, N11008\_K4096 (shapes determined by the Llama2-7B model), and N16384\_K16384, $M$ is increased from 4096 to 16384 in steps of 512, ensuring the size is a multiple of the TensorCore compute scale. These three curves demonstrate that blindly increasing the batch size does not always yield improved peak performance. Once the batch size is sufficiently large, further improvements in GEMM peak can be achieved by increasing $K$ and $N$. For the unaligned\_N11008\_K4096 case, our $M$ starts from 4096+13 (the magic number 13 is an odd number chosen to not significantly affect the size of $M$) and increases to 16384+13 in steps of 512. We analyze the performance differences between $M$ as integer multiples of the TensorCore compute scale and non-integer multiples. The results clearly show that when $M$ is an integer multiple of the TensorCore compute scale, the peak performance is higher than for non-integer multiples.
\vspace{-10px}
\begin{table}[!ht]
\centering
\caption{Performance comparison of the first GEMM in MLP with Naive and Recomputation methods. The data is from the average of 320 measurements.}
\begin{tabular}{c|c|c}
\hline
               & Naive            & Recomputation      \\ \hline
Shape($M$,$N$,$K$) & 666,11008,4096 & 10624,11008,4096 \\ \hline
Time(ms)       & 0.289            & 3.870              \\ \hline
Peak(\%)       & 66.6             & 79.4            \\ \hline
\end{tabular}
\label{table:GEMMPerf}
\vspace{-8px}
\end{table}

\begin{table}[H]
\centering
\caption{Time breakdown of GEMM kernels in forward and backward phases using naive and recomputation methods.}
\begin{tabular}{c|c|c}
\hline
              & Forward & Backward \\ \hline
Naive         & 66.4\%  & 62.5\%   \\ \hline
Recomputation & 66.1\%  & 69.0\%   \\ \hline
\end{tabular}
\label{table:Gemmproportion}
\end{table}

\subsection{Memory Copy} 
\begin{figure}[!ht]
     \centering
     \begin{subfigure}[b]{0.23\textwidth}
        \centering
        \includegraphics[width=\linewidth]{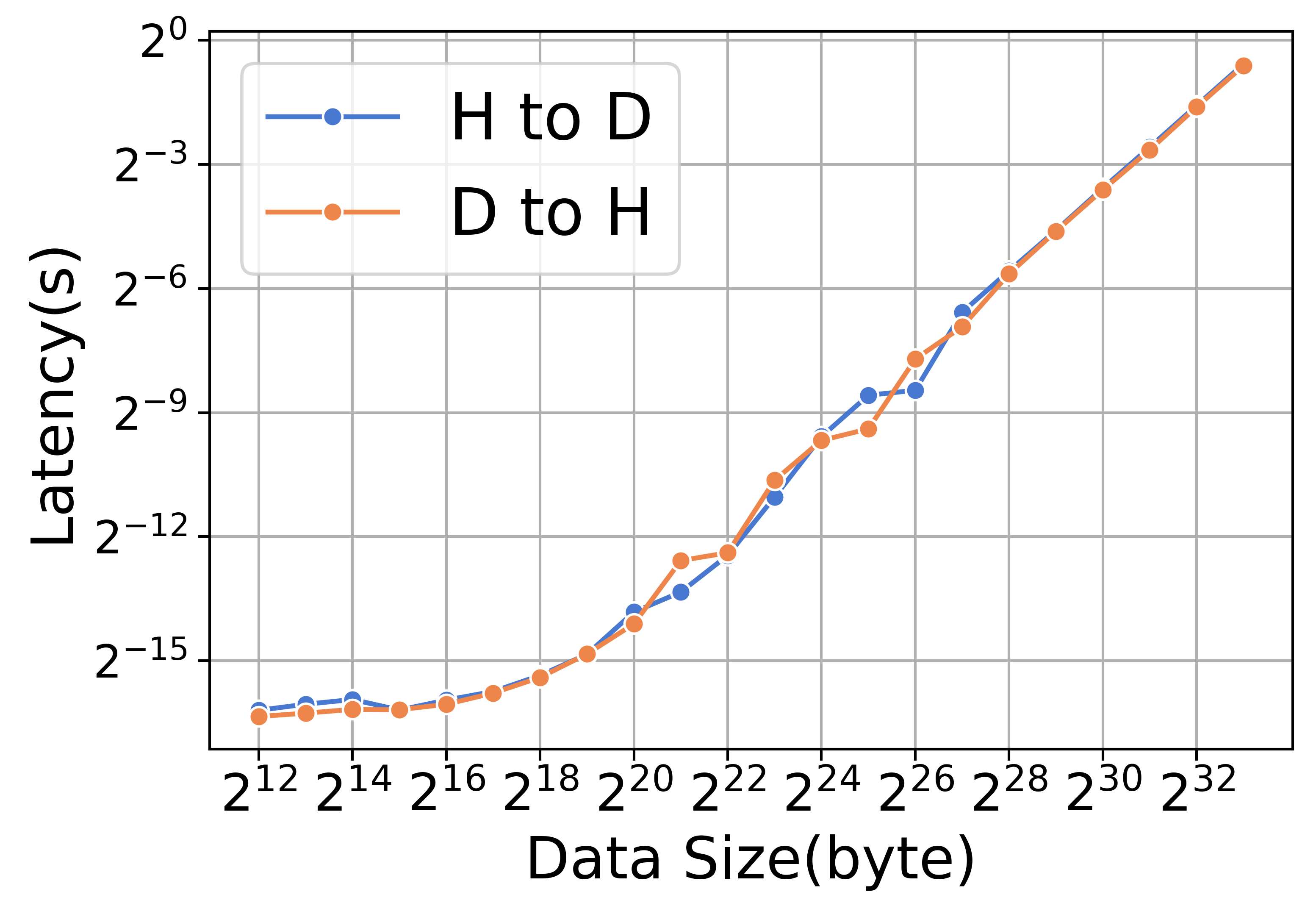}
        \caption*{(a) Latency}
        \label{fig:Memcpy-A800-L}
     \end{subfigure}
     \begin{subfigure}[b]{0.23\textwidth}
         \centering
        \includegraphics[width=\linewidth]{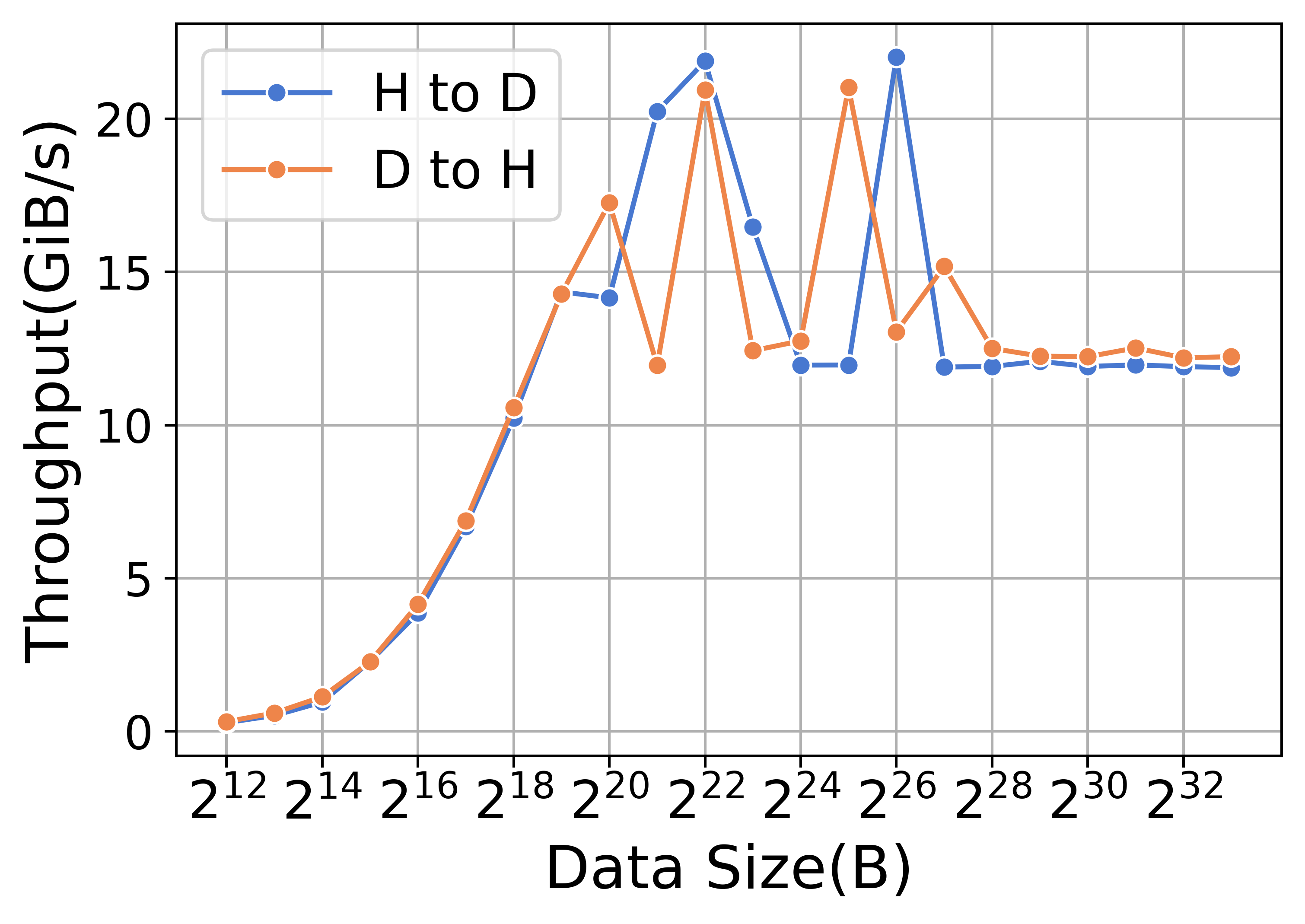}
        \caption*{(b) Throughput}
        \label{fig:Memcpy-A800-T}
     \end{subfigure}
        \caption{Latency and Throughput performance of offloading and uploading comparison on A800 with different data size.}
        \label{fig:Memcpy-A800}
\end{figure}

Offloading and uploading operations are implemented using memory copy kernels. Table~\ref{tab:memory-A800} summarizes the absolute time cost and the percentage of memory copy in each iteration on the A800 platform. As indicated in Table~\ref{tab:memory-A800}, ZeRO-3 incurs greater uploading and offloading times compared to ZeRO-2. However, the impact of memory copy in this context is relatively minor. Figure~\ref{fig:Memcpy-A800} demonstrates the performance of upload operations (denoted H to D) and offload operations (denoted D to H) in terms of kernels. This figure reveals that the throughput and latency for both uploading and offloading operations are similar. For smaller data sizes, the startup time tends to be dominant, while for larger data sizes, bandwidth becomes increasingly crucial.

\begin{table}[!ht]
    \centering
    \caption{We use the DeepSpeed framework, set (bf16) and batch size of 32. The absolute time and percentage of memory copy in each iteration in A800. The impact of memory copy in this setting is relatively minor.}
    \begin{tabular}{c|c|c|c} \hline  
         Method&  Model&   Time(s/iteration)& Percentage(\%)\\ \hline  
         \multirow{2}{*}{ZeRO-2}&  Llama2-7B&  0.596& 4.9\%\\ \cline{2-4}   
         &  Llama2-13B&  1.160& 7.3\%\\ \hline 
 \multirow{2}{*}{ZeRO-3}& Llama2-7B& 0.638&4.0\%\\ \cline{2-4}
 & Llama2-13B& 1.560&6.7\%\\ \hline
    \end{tabular}
    \label{tab:memory-A800}
\end{table}

\subsection{Collective Communication} 

\begin{figure}[!ht]
     \centering
     \begin{subfigure}[b]{0.23\textwidth}
        \centering
        \includegraphics[width=\linewidth]{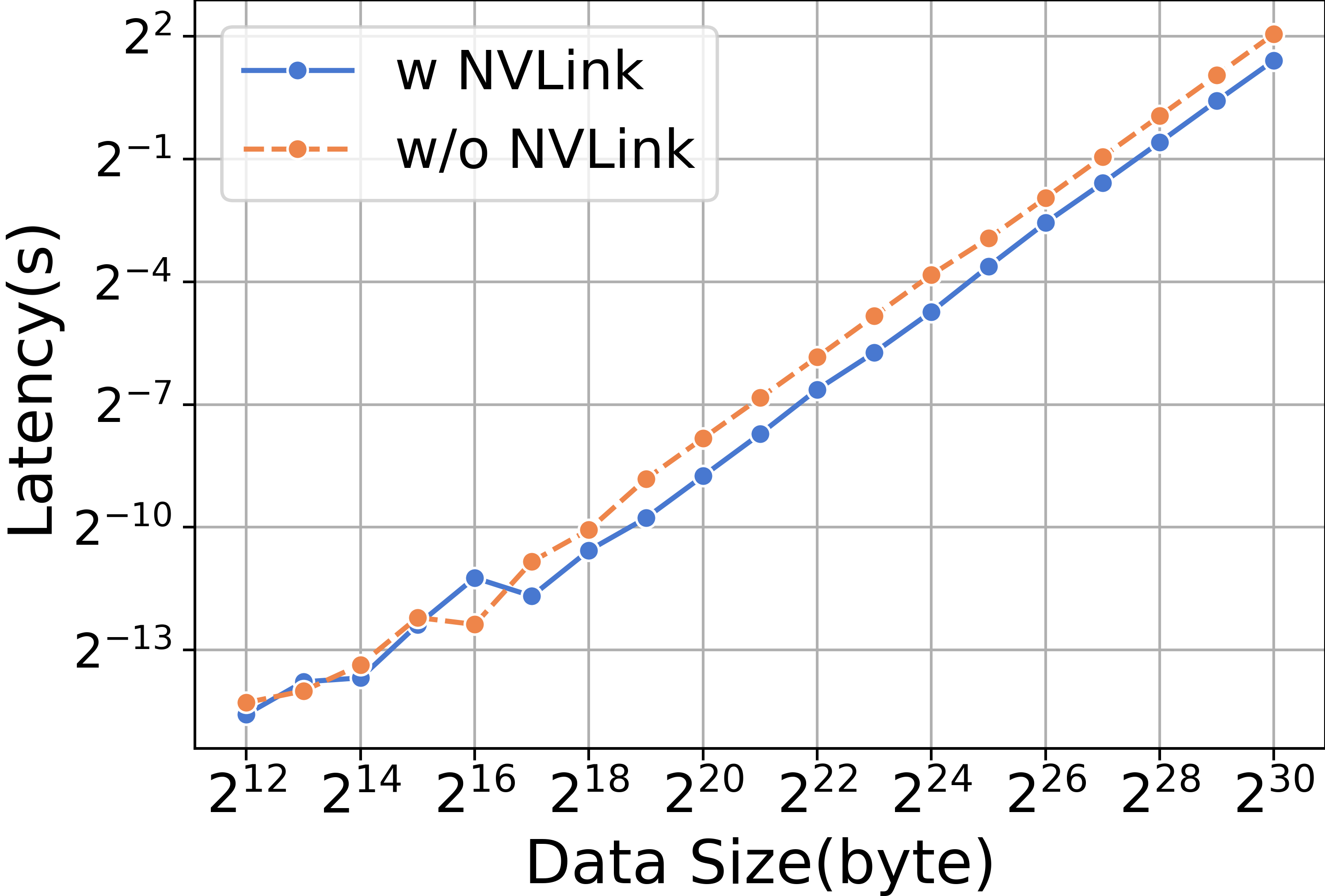}
        \caption*{(a) Latency}
        \label{fig:RTX-AllGather-L}
     \end{subfigure}
     \begin{subfigure}[b]{0.23\textwidth}
         \centering
        \includegraphics[width=\linewidth]{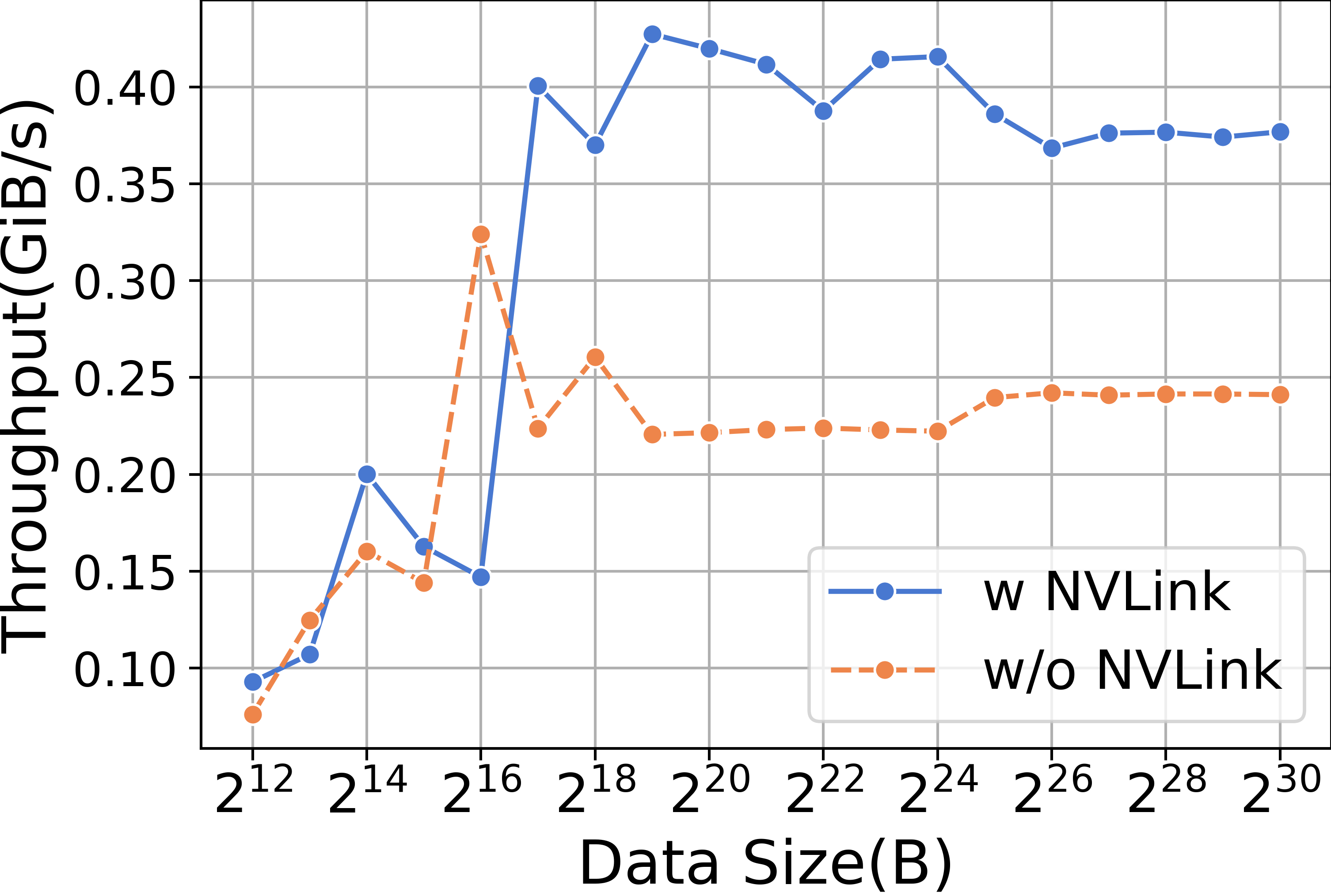}
        \caption*{(b) Throughput}
        \label{fig:RTX-AllGather-T}
     \end{subfigure}
     \vspace{-5px}
        \caption{The Latency and Throughput of AllGather on RTX3090 with and without NVLink with data size varied.}
        \label{fig:RTX-AllGather}
        \vspace{-5px}
\end{figure}

We first emphasize the high communication speed provided by NVLink. When conducting AllGather operations at various data scales, it is observed that the RTX3090 equipped with NVLink significantly outperforms its counterpart without NVLink, as illustrated in Figure~\ref{fig:RTX-AllGather}.

Different training paradigms involve various collective communication operations. In data parallel paradigms, AllReduce is used during the backward phase to synchronize weights, as illustrated in Table~\ref{tab:commuciation-ratio}.

\begin{figure}[!ht]
     \centering
     \begin{subfigure}[b]{0.23\textwidth}
        \centering
        \includegraphics[width=\linewidth]{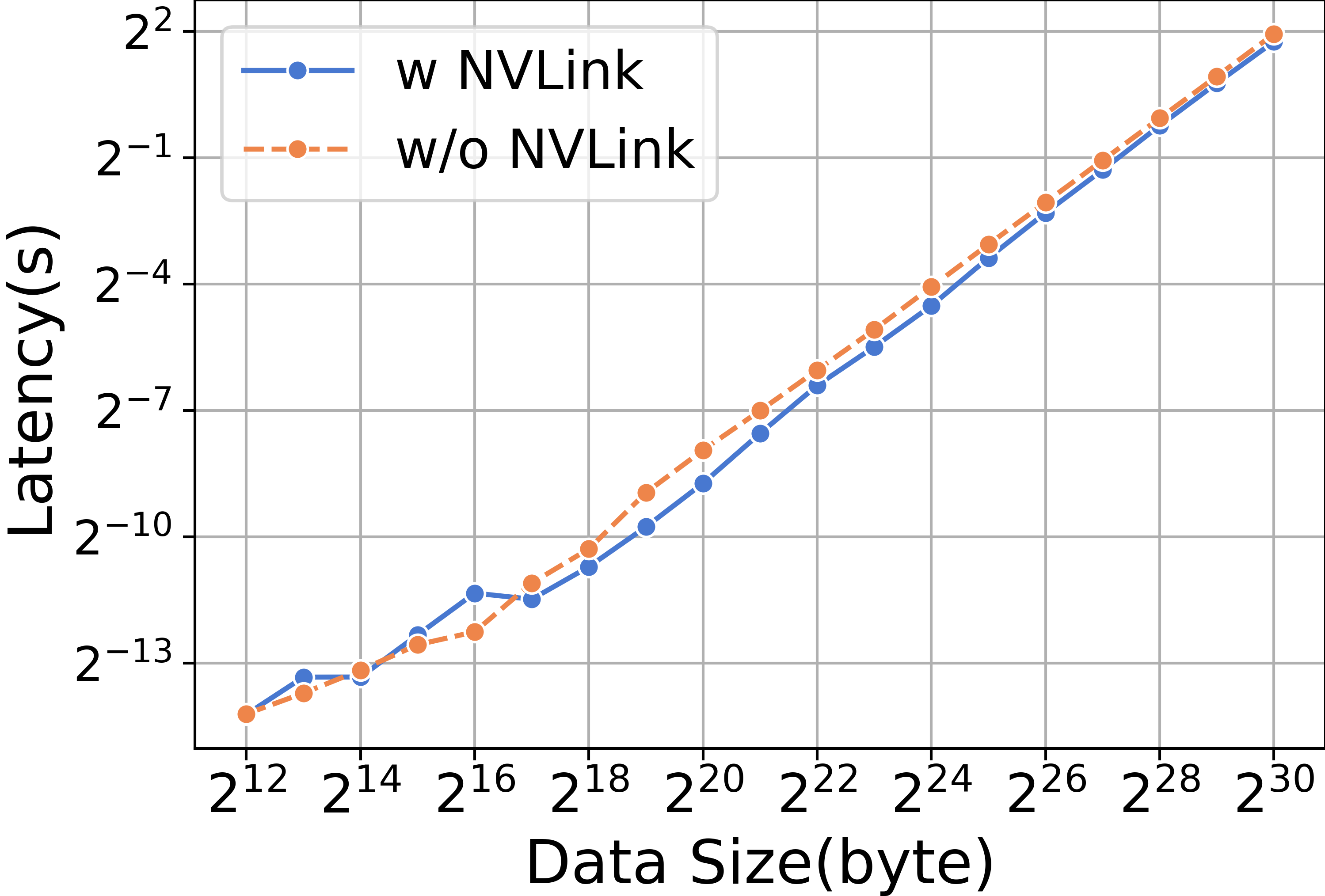}
        \caption*{(a) Latency}
        \label{fig:RTX-ReduceScatter-L}
     \end{subfigure}
     \begin{subfigure}[b]{0.23\textwidth}
        \centering
        \includegraphics[width=\linewidth]{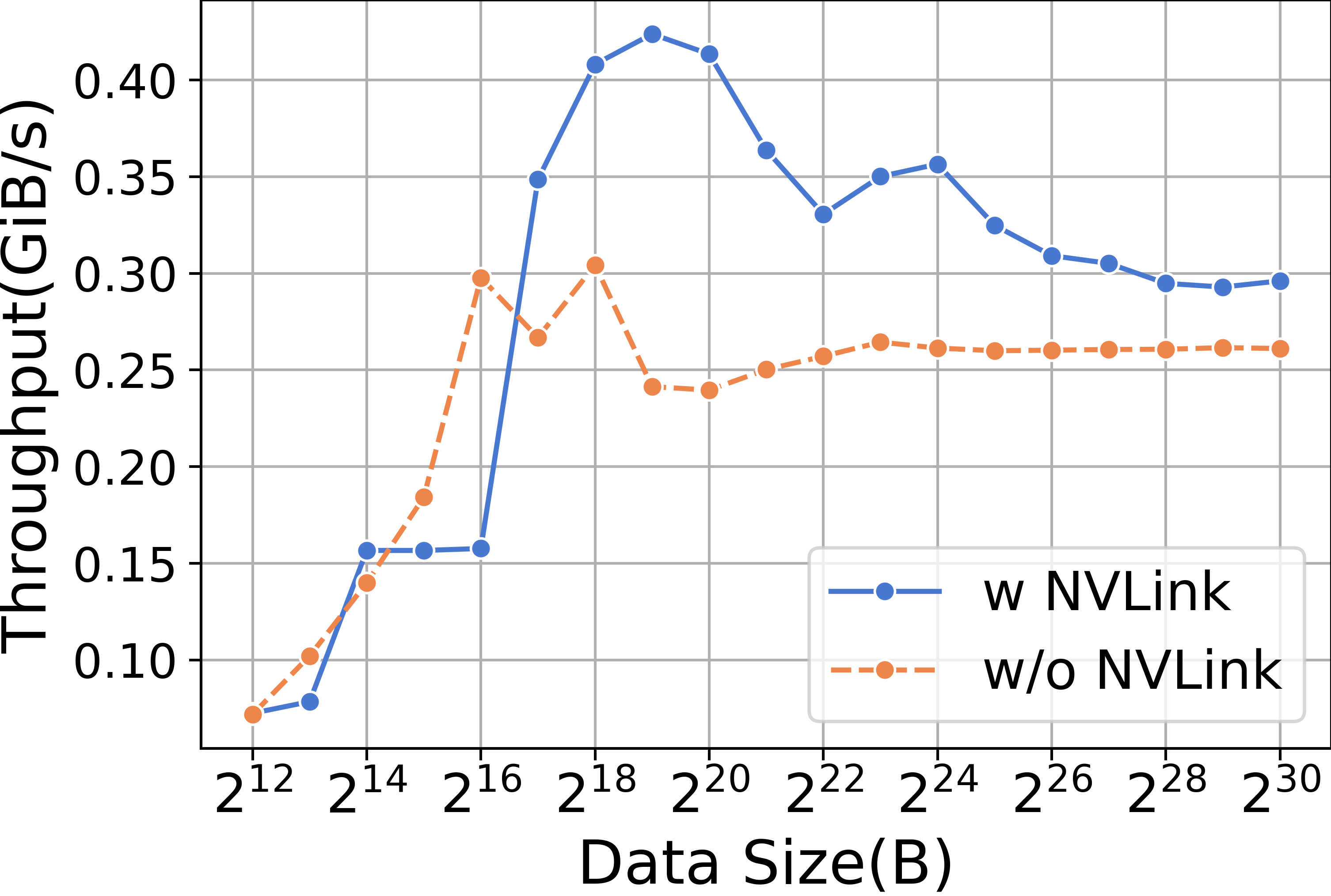}
        \caption*{(b) Throughput}
        \label{fig:RTX-ReduceScatter-T}
     \end{subfigure}
        \caption{The Latency and Throughput of ReduceScatter on RTX3090 with and without NVLink across different data size.}
        \label{fig:RTX-ReduceScatter}
\end{figure}

When conducting experiments with ReduceScatter and varying data sizes, it is observed that the RTX3090 with NVLink significantly outperforms its counterpart without NVLink, as demonstrated in Figure~\ref{fig:RTX-ReduceScatter}.

\begin{table}[!ht]
\vspace{-8px}
    \centering
    \caption{AllReduce Percentage.}
    \begin{tabular}{c|c|c} \hline  
         Llama2-7B & Time(s/iteration)& Percentage(\%)\\ \hline 
         Naive & 0.24 & 45.00 \\ \hline
         F & 0.23 & 44.97 \\ \hline
         R & 0.86 & 25.31 \\ \hline
         R+F & 0.69 & 20.41 \\ \hline 
    \end{tabular}
    \label{tab:commuciation-ratio}
    \vspace{-10px}
\end{table}

\begin{figure}[!ht]
     \centering
     \begin{subfigure}[b]{0.23\textwidth}
        \centering
        \includegraphics[width=\linewidth]{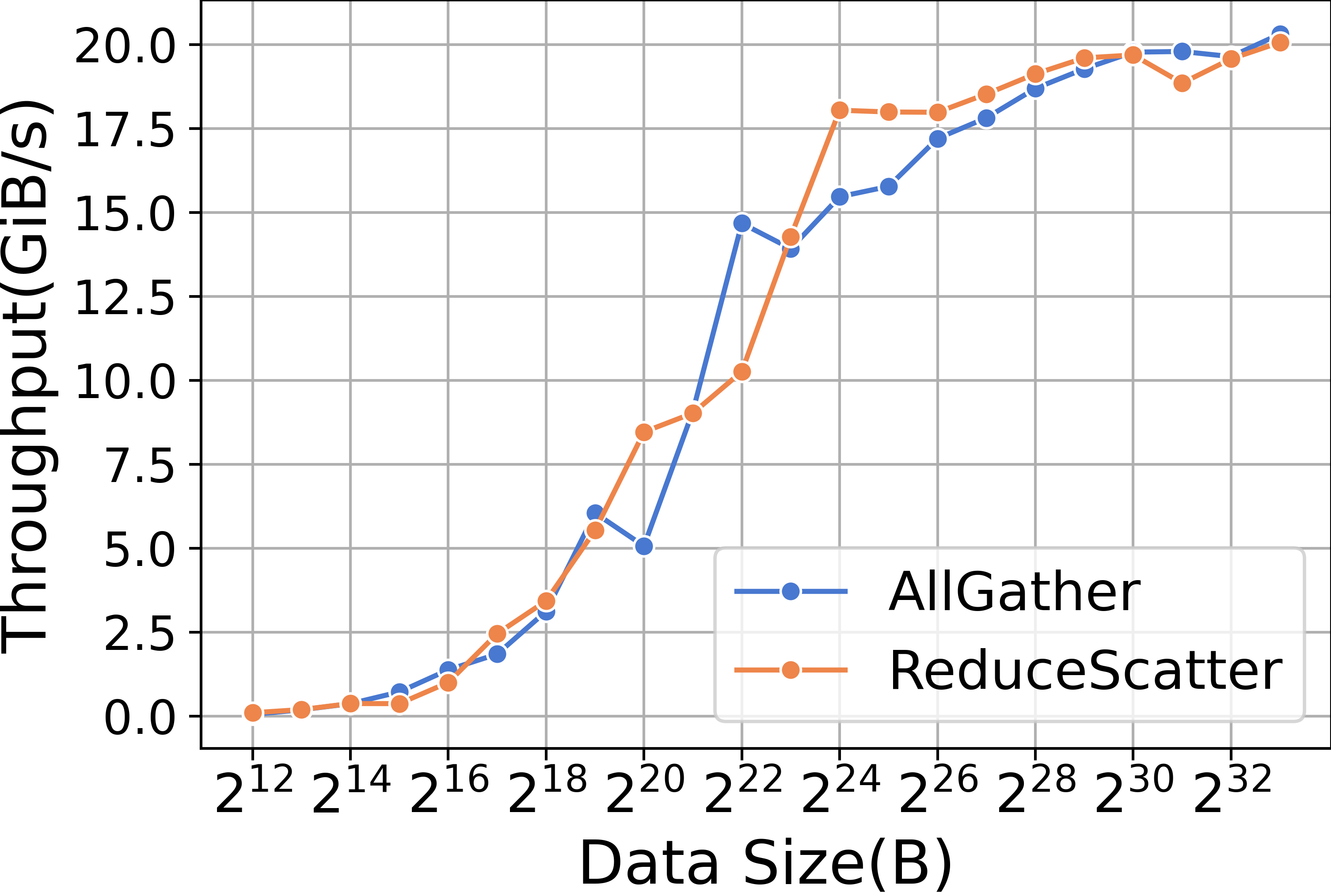}
        \phantomcaption
        \caption*{(a) AllGather, ReduceScatter}
        \label{fig:AllGather-A800-T}
     \end{subfigure}
     \begin{subfigure}[b]{0.23\textwidth}
         \centering
        \includegraphics[width=\linewidth]{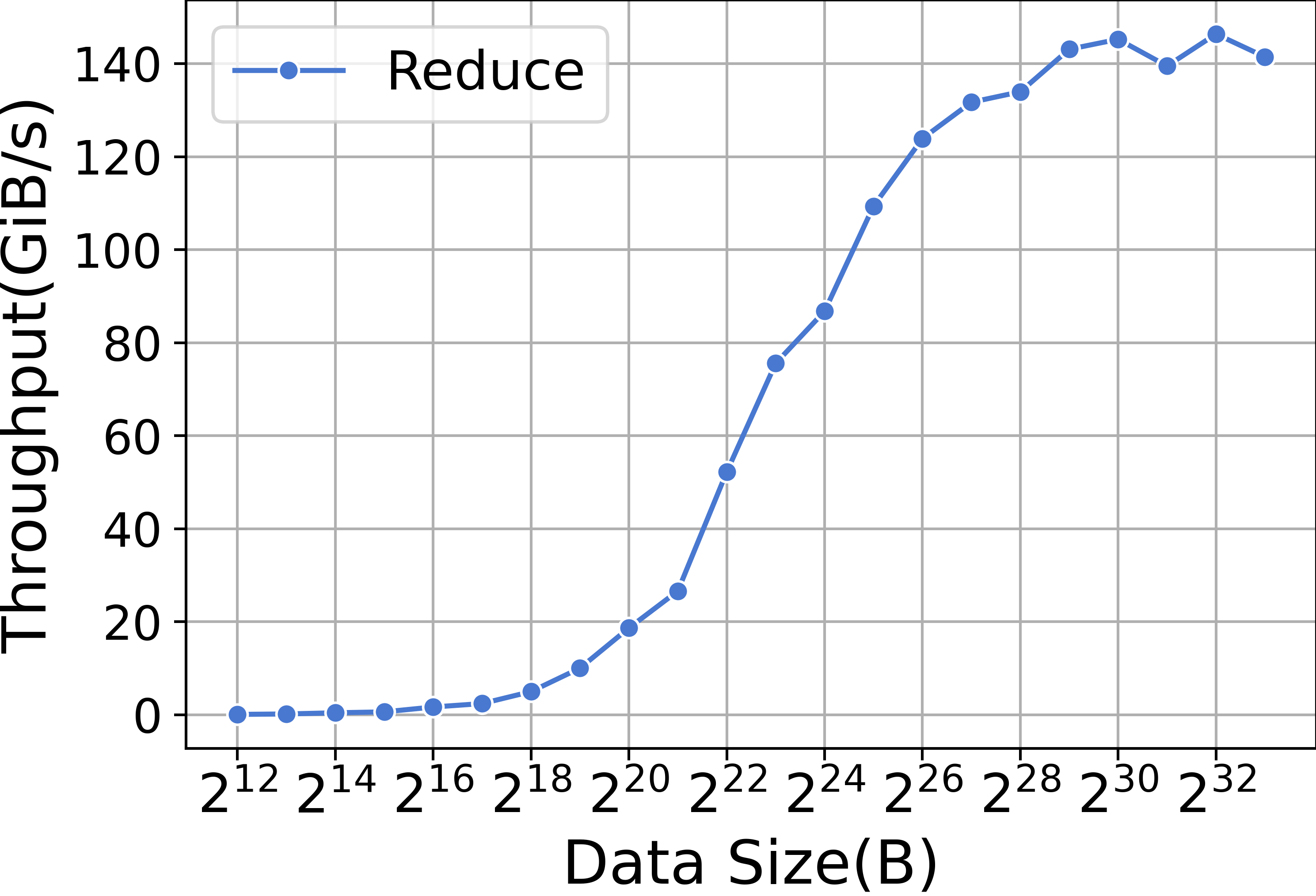}
        \phantomcaption
        \caption*{(b) Reduce}
        \label{fig:Reduce-A800-T}
     \end{subfigure}
        \caption{Throughput of AllGather, ReduceScatter, and Reduce on A800 with data size varied.}
        \label{fig:Reduce-A800}
        \vspace{-10px}
\end{figure}

ZeRO-2 requires the use of Reduce collective communication primitives in the backward phase. Figure~\ref{fig:Reduce-A800-T} displays the performance of Reduce kernels. Similar to memory copying, the small data size of the Reduce kernel results in the dominance of startup time, whereas the performance with large data sizes is dependent on bandwidth. In contrast, ZeRO-3 employs ReduceScatter instead of Reduce for collective communication in the backward phase. Figure~\ref{fig:AllGather-A800-T} illustrates the performance of the ReduceScatter kernels. Both ZeRO-2 and ZeRO-3 utilize AllGather for updating parameters, and Figure~\ref{fig:AllGather-A800-T} also presents the performance of AllGather kernels.

\begin{table}[!ht]
    \centering
    \caption{We use the DeepSpeed framework, set (bf16) and batch size of 32. The absolute time and percentage of communication kernel in each iteration in A800.}
    \begin{tabular}{c|c|c|c} \hline  
         Method&  Model&   Time(s/iteration)& Percentage(\%)\\ \hline  
         \multirow{2}{*}{ZeRO-2}&  Llama2-7B&  4.254& 41.8\%\\ \cline{2-4}   
         &  Llama2-13B&  3.779& 27.4\%\\ \hline 
 \multirow{2}{*}{ZeRO-3}& Llama2-7B& 4.576&28.1\%\\ \cline{2-4}
 & Llama2-13B& 2.791&11.9\%\\ \hline
    \end{tabular}
    \label{tab:comm-A800}
    \vspace{-10px}
\end{table}
Table~\ref{tab:comm-A800} summarizes the absolute time and percentage of memory copy in each iteration in A800. As shown in Table~\ref{tab:comm-A800}, we can see that ZeRO-3 has more communication time than ZeRO-2, and communication time is more significant in pre-training small models than larger models.


\section{Related Works}\label{sec:relatedworks}
Extensive studies have benchmarked model performance in terms of generalization capability and accuracy in downstream tasks\cite{zhao2023survey,kaddour2023challenges,fu2023mme,valmeekam2023planning,chang2023survey}. However, few studies focus on evaluating and analyzing time to study hardware\cite{9139681,9139835} and software\cite{xu2017performance,9275615} efficiency, and even fewer address these aspects in training, fine-tuning, and serving LLMs. AIPerf\cite{ren2021aiperf} implement the algorithms in a highly parallel and flexible way, and evaluate their performance on various systems. MLPerf\cite{reddi2019mlperf} is another cutting-edge benchmark for comparing the time performance of deep learning (which includes LLMs) in training and inference, without limitations on hardware platforms. In the pre-LLM era, benchmarks were proposed to compare software and hardware performance across various models, including CNNs, LSTMs, and Transformers~\cite{shi2016benchmarking,shi2018performance,li2020characterizing,tang2020communication,jansen2020ddlbench,liang2022benchmark,lu2023quantitative}. Xu et al.\cite{xu2023survey} provided a survey comparing model compression techniques, which are particularly useful in model fine-tuning and inference. Cao et al.\cite{cao2023comprehensive} also provided an overview of efficient LLMs, focusing on algorithmic aspects such as ELECTRA~\cite{Clark2020ELECTRA}, Prompt Tuning, etc. Liang et al.,~\cite{liang2023holistic} proposed HELM, a comprehensive performance evaluation, for comparing both model generalization capability and time efficiency. However, their results on time efficiency focus on inference using a specific hardware platform with given software. Specifically in inference, LLMPerf\cite{llmperf} benchmarks the throughput performance of various LLMs.

To the best of our knowledge, this is the first study to analyze runtime performance across all three key stages (pre-training, fine-tuning, and serving) of LLMs on various hardware platforms.

\section{Conclusion}\label{sec:conclusion}
In this work, we have benchmarked the runtime performance of pre-training, fine-tuning, and serving LLMs on three 8-GPU hardware platforms: Nvidia A800-80G, RTX4090, and RTX3090. Based on the benchmark results, we analyzed the key modules and operators that are major contributors to the overall time. The experimental results and analyses provide more information for end-users in choosing configurations in terms of hardware, software, and optimization techniques for pre-training, fine-tuning, and serving LLMs. Additionally, an in-depth understanding of performance offers further opportunities for system optimization.

\section*{Acknowledgement}

We would like to thank Mr. Yongke Zhao for the valuable feedback. This work was partially supported by National Natural Science Foundation of China under Grant No. 62272122, a Hong Kong RIF grant under Grant No. R6021-20, and Hong Kong CRF grants under Grant No. C2004-21GF and C7004-22GF. We also thank the HPC-AI-Integrated Intelligent Computing center of HKUST(GZ) for providing some of the hardware platforms in this project. 
\section*{Revision history}

\textbf{Wed, 29 Nov 2023}
\begin{itemize}
   \item We corrected the throughput data in Fig. \ref{fig:scaling} and Tables \ref{table:ds-vs-meg}, \ref{apx-table:many-method-bs1}, \ref{apx-table:many-method-bs32} and \ref{apx-table:finetuning-manymethod-bs1}. The mistakes were caused by a bug in the code that calculates the number of sequence tokens.
   \item We updated the detailed introduction to benchmark settings, including the datasets, sequence length, etc., in Sec. \ref{sec:methodologies}.
\end{itemize}

\bibliography{cites}

\begin{thebibliography}{10}
\providecommand{\url}[1]{#1}
\csname url@samestyle\endcsname
\providecommand{\newblock}{\relax}
\providecommand{\bibinfo}[2]{#2}
\providecommand{\BIBentrySTDinterwordspacing}{\spaceskip=0pt\relax}
\providecommand{\BIBentryALTinterwordstretchfactor}{4}
\providecommand{\BIBentryALTinterwordspacing}{\spaceskip=\fontdimen2\font plus
\BIBentryALTinterwordstretchfactor\fontdimen3\font minus \fontdimen4\font\relax}
\providecommand{\BIBforeignlanguage}[2]{{%
\expandafter\ifx\csname l@#1\endcsname\relax
\typeout{** WARNING: IEEEtran.bst: No hyphenation pattern has been}%
\typeout{** loaded for the language `#1'. Using the pattern for}%
\typeout{** the default language instead.}%
\else
\language=\csname l@#1\endcsname
\fi
#2}}
\providecommand{\BIBdecl}{\relax}
\BIBdecl

\bibitem{ouyang2022training}
L.~Ouyang, J.~Wu, X.~Jiang, D.~Almeida, C.~Wainwright, P.~Mishkin, C.~Zhang, S.~Agarwal, K.~Slama, A.~Ray \emph{et~al.}, ``Training language models to follow instructions with human feedback,'' \emph{Advances in Neural Information Processing Systems}, vol.~35, pp. 27\,730--27\,744, 2022.

\bibitem{chang2023survey}
Y.~Chang, X.~Wang, J.~Wang, Y.~Wu, K.~Zhu, H.~Chen, L.~Yang, X.~Yi, C.~Wang, Y.~Wang \emph{et~al.}, ``A survey on evaluation of large language models,'' \emph{arXiv preprint arXiv:2307.03109}, 2023.

\bibitem{kaplan2020scaling}
J.~Kaplan, S.~McCandlish, T.~Henighan, T.~B. Brown, B.~Chess, R.~Child, S.~Gray, A.~Radford, J.~Wu, and D.~Amodei, ``Scaling laws for neural language models,'' \emph{arXiv preprint arXiv:2001.08361}, 2020.

\bibitem{hoffmann2022training}
J.~Hoffmann, S.~Borgeaud, A.~Mensch, E.~Buchatskaya, T.~Cai, E.~Rutherford, D.~d.~L. Casas, L.~A. Hendricks, J.~Welbl, A.~Clark \emph{et~al.}, ``Training compute-optimal large language models,'' \emph{arXiv preprint arXiv:2203.15556}, 2022.

\bibitem{openai2023gpt4}
OpenAI, ``{GPT}-4 technical report,'' 2023.

\bibitem{brown2020language}
T.~Brown, B.~Mann, N.~Ryder, M.~Subbiah, J.~D. Kaplan, P.~Dhariwal, A.~Neelakantan, P.~Shyam, G.~Sastry, A.~Askell \emph{et~al.}, ``Language models are few-shot learners,'' \emph{Advances in Neural Information Processing Systems}, vol.~33, pp. 1877--1901, 2020.

\bibitem{chowdhery2022palm}
A.~Chowdhery, S.~Narang, J.~Devlin, M.~Bosma, G.~Mishra, A.~Roberts, P.~Barham, H.~W. Chung, C.~Sutton, S.~Gehrmann \emph{et~al.}, ``{PaLM}: Scaling language modeling with pathways,'' in \emph{Proceedings of Machine Learning and Systems 2022}, 2022.

\bibitem{kaddour2023challenges}
J.~Kaddour, J.~Harris, M.~Mozes, H.~Bradley, R.~Raileanu, and R.~McHardy, ``Challenges and applications of large language models,'' \emph{arXiv preprint arXiv:2307.10169}, 2023.

\bibitem{touvron2023llama_2}
H.~Touvron, L.~Martin, K.~Stone, P.~Albert, A.~Almahairi, Y.~Babaei, N.~Bashlykov, S.~Batra, P.~Bhargava, S.~Bhosale \emph{et~al.}, ``Llama 2: Open foundation and fine-tuned chat models,'' \emph{arXiv preprint arXiv:2307.09288}, 2023.

\bibitem{rasley2020deepspeed}
J.~Rasley, S.~Rajbhandari, O.~Ruwase, and Y.~He, ``{DeepSpeed}: System optimizations enable training deep learning models with over 100 billion parameters,'' in \emph{Proceedings of the 26th ACM SIGKDD International Conference on Knowledge Discovery \& Data Mining}, 2020, pp. 3505--3506.

\bibitem{narayanan2021efficient}
D.~Narayanan, M.~Shoeybi, J.~Casper, P.~LeGresley, M.~Patwary, V.~Korthikanti, D.~Vainbrand, P.~Kashinkunti, J.~Bernauer, B.~Catanzaro \emph{et~al.}, ``Efficient large-scale language model training on gpu clusters using megatron-lm,'' in \emph{Proceedings of the International Conference for High Performance Computing, Networking, Storage and Analysis}, 2021, pp. 1--15.

\bibitem{peft}
S.~Mangrulkar, S.~Gugger, L.~Debut, Y.~Belkada, S.~Paul, and B.~Bossan, ``{PEFT}: State-of-the-art parameter-efficient fine-tuning methods,'' \url{https://github.com/huggingface/peft}, 2022.

\bibitem{kwon2023efficient}
W.~Kwon, Z.~Li, S.~Zhuang, Y.~Sheng, L.~Zheng, C.~H. Yu, J.~E. Gonzalez, H.~Zhang, and I.~Stoica, ``Efficient memory management for large language model serving with pagedattention,'' in \emph{Proceedings of the ACM SIGOPS 29th Symposium on Operating Systems Principles}, 2023.

\bibitem{LightLLM2023}
GitHub, ``{LightLLM}: A python-based large language model inference and serving framework,'' \url{https://github.com/ModelTC/lightllm}, 2023.

\bibitem{TGI2023}
HuggingFace, ``Text generation inference,'' \url{https://github.com/huggingface/text-generation-inference}, 2023.

\bibitem{rajbhandari2020zero}
S.~Rajbhandari, J.~Rasley, O.~Ruwase, and Y.~He, ``{ZeRO}: Memory optimizations toward training trillion parameter models,'' in \emph{SC20: International Conference for High Performance Computing, Networking, Storage and Analysis}.\hskip 1em plus 0.5em minus 0.4em\relax IEEE, 2020, pp. 1--16.

\bibitem{korthikanti2023reducing}
V.~A. Korthikanti, J.~Casper, S.~Lym, L.~McAfee, M.~Andersch, M.~Shoeybi, and B.~Catanzaro, ``Reducing activation recomputation in large transformer models,'' \emph{Proceedings of Machine Learning and Systems}, vol.~5, 2023.

\bibitem{jain2020checkmate}
P.~Jain, A.~Jain, A.~Nrusimha, A.~Gholami, P.~Abbeel, K.~Keutzer, I.~Stoica, and J.~E. Gonzalez, ``Checkmate: Breaking the memory wall with optimal tensor rematerialization,'' \emph{arXiv preprint arXiv:1910.02653}, 2020.

\bibitem{smith2022using}
S.~Smith, M.~Patwary, B.~Norick, P.~LeGresley, S.~Rajbhandari, J.~Casper, Z.~Liu, S.~Prabhumoye, G.~Zerveas, V.~Korthikanti, E.~Zhang, R.~Child, R.~Y. Aminabadi, J.~Bernauer, X.~Song, M.~Shoeybi, Y.~He, M.~Houston, S.~Tiwary, and B.~Catanzaro, ``Using {DeepSpeed} and {Megatron} to train {Megatron}-turing {NLG} 530{B}, a large-scale generative language model,'' \emph{arXiv preprint arXiv:2201.11990}, 2022.

\bibitem{yao2022zeroquant}
Z.~Yao, R.~Yazdani~Aminabadi, M.~Zhang, X.~Wu, C.~Li, and Y.~He, ``Zeroquant: Efficient and affordable post-training quantization for large-scale transformers,'' \emph{Advances in Neural Information Processing Systems}, vol.~35, pp. 27\,168--27\,183, 2022.

\bibitem{hu2022lora}
\BIBentryALTinterwordspacing
E.~J. Hu, yelong shen, P.~Wallis, Z.~Allen-Zhu, Y.~Li, S.~Wang, L.~Wang, and W.~Chen, ``Lo{RA}: Low-rank adaptation of large language models,'' in \emph{International Conference on Learning Representations}, 2022. [Online]. Available: \url{https://openreview.net/forum?id=nZeVKeeFYf9}
\BIBentrySTDinterwordspacing

\bibitem{dettmers2023qlora}
T.~Dettmers, A.~Pagnoni, A.~Holtzman, and L.~Zettlemoyer, ``{QLoRA}: Efficient finetuning of quantized {LLMs},'' \emph{arXiv preprint arXiv:2305.14314}, 2023.

\bibitem{dao2022flashattention}
T.~Dao, D.~Fu, S.~Ermon, A.~Rudra, and C.~R{\'e}, ``Flashattention: Fast and memory-efficient exact attention with io-awareness,'' \emph{Advances in Neural Information Processing Systems}, vol.~35, pp. 16\,344--16\,359, 2022.

\bibitem{vaswani2017attention}
A.~Vaswani, N.~Shazeer, N.~Parmar, J.~Uszkoreit, L.~Jones, A.~N. Gomez, {\L}.~Kaiser, and I.~Polosukhin, ``Attention is all you need,'' \emph{Advances in Neural Information Processing Systems}, vol.~30, 2017.

\bibitem{GPT3}
\BIBentryALTinterwordspacing
T.~Brown, B.~Mann, N.~Ryder, M.~Subbiah, J.~D. Kaplan, P.~Dhariwal, A.~Neelakantan, P.~Shyam, G.~Sastry, A.~Askell, S.~Agarwal, A.~Herbert-Voss, G.~Krueger, T.~Henighan, R.~Child, A.~Ramesh, D.~Ziegler, J.~Wu, C.~Winter, C.~Hesse, M.~Chen, E.~Sigler, M.~Litwin, S.~Gray, B.~Chess, J.~Clark, C.~Berner, S.~McCandlish, A.~Radford, I.~Sutskever, and D.~Amodei, ``Language models are {Few-Shot} learners,'' in \emph{Advances in Neural Information Processing Systems}, H.~Larochelle, M.~Ranzato, R.~Hadsell, M.~Balcan, and H.~Lin, Eds., vol.~33.\hskip 1em plus 0.5em minus 0.4em\relax Curran Associates, Inc., 2020, pp. 1877--1901. [Online]. Available: \url{https://proceedings.neurips.cc/paper_files/paper/2020/file/1457c0d6bfcb4967418bfb8ac142f64a-Paper.pdf}
\BIBentrySTDinterwordspacing

\bibitem{touvron2023llama_1}
H.~Touvron, T.~Lavril, G.~Izacard, X.~Martinet, M.-A. Lachaux, T.~Lacroix, B.~Rozière, N.~Goyal, E.~Hambro, F.~Azhar, A.~Rodriguez, A.~Joulin, E.~Grave, and G.~Lample, ``{LLaMA}: Open and efficient foundation language models,'' \emph{arXiv preprint arXiv:2302.13971}, 2023.

\bibitem{scao2022bloom}
T.~L. Scao, A.~Fan, C.~Akiki, E.~Pavlick, S.~Ili{\'c}, D.~Hesslow, R.~Castagn{\'e}, A.~S. Luccioni, F.~Yvon, M.~Gall{\'e} \emph{et~al.}, ``Bloom: A 176b-parameter open-access multilingual language model,'' \emph{arXiv preprint arXiv:2211.05100}, 2022.

\bibitem{ren2021zerooffload}
J.~Ren, S.~Rajbhandari, R.~Y. Aminabadi, O.~Ruwase, S.~Yang, M.~Zhang, D.~Li, and Y.~He, ``{ZeRO-Offload}: Democratizing {Billion-Scale} model training,'' in \emph{2021 USENIX Annual Technical Conference (USENIX ATC 21)}, 2021, pp. 551--564.

\bibitem{rajbhandari2021zeroinfinity}
S.~Rajbhandari, O.~Ruwase, J.~Rasley, S.~Smith, and Y.~He, ``Zero-infinity: Breaking the gpu memory wall for extreme scale deep learning,'' in \emph{Proceedings of the International Conference for High Performance Computing, Networking, Storage and Analysis}, 2021, pp. 1--14.

\bibitem{aminabadi2022deepspeed}
R.~Y. Aminabadi, S.~Rajbhandari, A.~A. Awan, C.~Li, D.~Li, E.~Zheng, O.~Ruwase, S.~Smith, M.~Zhang, J.~Rasley \emph{et~al.}, ``{DeepSpeed}-inference: enabling efficient inference of transformer models at unprecedented scale,'' in \emph{SC22: International Conference for High Performance Computing, Networking, Storage and Analysis}.\hskip 1em plus 0.5em minus 0.4em\relax IEEE, 2022, pp. 1--15.

\bibitem{shoeybi2019megatron}
M.~Shoeybi, M.~Patwary, R.~Puri, P.~LeGresley, J.~Casper, and B.~Catanzaro, ``Megatron-{LM}: Training multi-billion parameter language models using model parallelism,'' \emph{arXiv preprint arXiv:1909.08053}, 2019.

\bibitem{lester2021power}
B.~Lester, R.~Al-Rfou, and N.~Constant, ``The power of scale for parameter-efficient prompt tuning,'' in \emph{Proceedings of the 2021 Conference on Empirical Methods in Natural Language Processing}, 2021, pp. 3045--3059.

\bibitem{dean2012large}
J.~Dean, G.~Corrado, R.~Monga, K.~Chen, M.~Devin, M.~Mao, M.~Ranzato, A.~Senior, P.~Tucker, K.~Yang \emph{et~al.}, ``Large scale distributed deep networks,'' \emph{Advances in Neural Information Processing Systems}, vol.~25, 2012.

\bibitem{vinoski2012server}
S.~Vinoski, ``Server-sent events with yaws,'' \emph{IEEE internet computing}, vol.~16, no.~5, pp. 98--102, 2012.

\bibitem{elfwing2018sigmoid}
S.~Elfwing, E.~Uchibe, and K.~Doya, ``Sigmoid-weighted linear units for neural network function approximation in reinforcement learning,'' \emph{Neural Networks}, vol. 107, pp. 3--11, 2018.

\bibitem{zhang2019root}
B.~Zhang and R.~Sennrich, ``Root mean square layer normalization,'' \emph{Advances in Neural Information Processing Systems}, vol.~32, 2019.

\bibitem{zhao2023survey}
W.~X. Zhao, K.~Zhou, J.~Li, T.~Tang, X.~Wang, Y.~Hou, Y.~Min, B.~Zhang, J.~Zhang, Z.~Dong \emph{et~al.}, ``A survey of large language models,'' \emph{arXiv preprint arXiv:2303.18223}, 2023.

\bibitem{fu2023mme}
C.~Fu, P.~Chen, Y.~Shen, Y.~Qin, M.~Zhang, X.~Lin, Z.~Qiu, W.~Lin, J.~Yang, X.~Zheng \emph{et~al.}, ``{MME}: A comprehensive evaluation benchmark for multimodal large language models,'' \emph{arXiv preprint arXiv:2306.13394}, 2023.

\bibitem{valmeekam2023planning}
K.~Valmeekam, S.~Sreedharan, M.~Marquez, A.~Olmo, and S.~Kambhampati, ``On the planning abilities of large language models (a critical investigation with a proposed benchmark),'' \emph{arXiv preprint arXiv:2302.06706}, 2023.

\bibitem{9139681}
Y.~Wang, Q.~Wang, S.~Shi, X.~He, Z.~Tang, K.~Zhao, and X.~Chu, ``Benchmarking the performance and energy efficiency of {AI} accelerators for {AI} training,'' in \emph{2020 20th IEEE/ACM International Symposium on Cluster, Cloud and Internet Computing (CCGRID)}, 2020, pp. 744--751.

\bibitem{9139835}
D.~Yan, W.~Wang, and X.~Chu, ``Demystifying tensor cores to optimize half-precision matrix multiply,'' in \emph{2020 IEEE International Parallel and Distributed Processing Symposium (IPDPS)}, 2020, pp. 634--643.

\bibitem{xu2017performance}
P.~Xu, S.~Shi, and X.~Chu, ``Performance evaluation of deep learning tools in docker containers,'' in \emph{2017 3rd International Conference on Big Data Computing and Communications (BIGCOM)}.\hskip 1em plus 0.5em minus 0.4em\relax IEEE, 2017, pp. 395--403.

\bibitem{9275615}
S.~Shi, Z.~Tang, X.~Chu, C.~Liu, W.~Wang, and B.~Li, ``{A Quantitative Survey of Communication Optimizations in Distributed Deep Learning},'' \emph{IEEE Network}, vol.~35, no.~3, pp. 230--237, 2021.

\bibitem{ren2021aiperf}
Z.~Ren, Y.~Liu, T.~Shi, L.~Xie, Y.~Zhou, J.~Zhai, Y.~Zhang, Y.~Zhang, and W.~Chen, ``{AIPerf: Automated machine learning as an AI-HPC benchmark},'' 2021.

\bibitem{reddi2019mlperf}
V.~J. Reddi, C.~Cheng, D.~Kanter, P.~Mattson, G.~Schmuelling, C.-J. Wu, B.~Anderson, M.~Breughe, M.~Charlebois, W.~Chou, R.~Chukka, C.~Coleman, S.~Davis, P.~Deng, G.~Diamos, J.~Duke, D.~Fick, J.~S. Gardner, I.~Hubara, S.~Idgunji, T.~B. Jablin, J.~Jiao, T.~S. John, P.~Kanwar, D.~Lee, J.~Liao, A.~Lokhmotov, F.~Massa, P.~Meng, P.~Micikevicius, C.~Osborne, G.~Pekhimenko, A.~T.~R. Rajan, D.~Sequeira, A.~Sirasao, F.~Sun, H.~Tang, M.~Thomson, F.~Wei, E.~Wu, L.~Xu, K.~Yamada, B.~Yu, G.~Yuan, A.~Zhong, P.~Zhang, and Y.~Zhou, ``{MLPerf Inference Benchmark},'' 2019.

\bibitem{shi2016benchmarking}
S.~Shi, Q.~Wang, P.~Xu, and X.~Chu, ``Benchmarking state-of-the-art deep learning software tools,'' in \emph{2016 7th International Conference on Cloud Computing and Big Data (CCBD)}.\hskip 1em plus 0.5em minus 0.4em\relax IEEE, 2016, pp. 99--104.

\bibitem{shi2018performance}
S.~Shi, Q.~Wang, and X.~Chu, ``Performance modeling and evaluation of distributed deep learning frameworks on {GPUs},'' in \emph{2018 IEEE 16th Intl Conf on Dependable, Autonomic and Secure Computing, 16th Intl Conf on Pervasive Intelligence and Computing, 4th Intl Conf on Big Data Intelligence and Computing and Cyber Science and Technology Congress (DASC/PiCom/DataCom/CyberSciTech)}.\hskip 1em plus 0.5em minus 0.4em\relax IEEE, 2018, pp. 949--957.

\bibitem{li2020characterizing}
S.~Li, R.~J. Walls, and T.~Guo, ``Characterizing and modeling distributed training with transient cloud {GPU} servers,'' in \emph{2020 IEEE 40th International Conference on Distributed Computing Systems (ICDCS)}.\hskip 1em plus 0.5em minus 0.4em\relax IEEE, 2020, pp. 943--953.

\bibitem{tang2020communication}
Z.~Tang, S.~Shi, X.~Chu, W.~Wang, and B.~Li, ``Communication-efficient distributed deep learning: A comprehensive survey,'' \emph{arXiv preprint arXiv:2003.06307}, 2020.

\bibitem{jansen2020ddlbench}
M.~Jansen, V.~Codreanu, and A.-L. Varbanescu, ``{DDLBench}: towards a scalable benchmarking infrastructure for distributed deep learning,'' in \emph{2020 IEEE/ACM Fourth Workshop on Deep Learning on Supercomputers (DLS)}.\hskip 1em plus 0.5em minus 0.4em\relax IEEE, 2020, pp. 31--39.

\bibitem{liang2022benchmark}
G.~Liang and I.~Alsmadi, ``Benchmark assessment for deepspeed optimization library,'' \emph{arXiv preprint arXiv:2202.12831}, 2022.

\bibitem{lu2023quantitative}
Z.~Lu, C.~Du, Y.~Jiang, X.~Xie, T.~Li, and F.~Yang, ``Quantitative evaluation of deep learning frameworks in heterogeneous computing environment,'' \emph{CCF Transactions on High Performance Computing}, pp. 1--18, 2023.

\bibitem{xu2023survey}
C.~Xu and J.~McAuley, ``A survey on model compression and acceleration for pretrained language models,'' in \emph{Proceedings of the AAAI Conference on Artificial Intelligence}, vol.~37, no.~9, 2023, pp. 10\,566--10\,575.

\bibitem{cao2023comprehensive}
Y.~Cao, S.~Li, Y.~Liu, Z.~Yan, Y.~Dai, P.~S. Yu, and L.~Sun, ``A comprehensive survey of {AI}-generated content ({AIGC}): A history of generative {AI} from {GAN} to {ChatGPT},'' \emph{arXiv preprint arXiv:2303.04226}, 2023.

\bibitem{Clark2020ELECTRA}
\BIBentryALTinterwordspacing
K.~Clark, M.-T. Luong, Q.~V. Le, and C.~D. Manning, ``{ELECTRA}: Pre-training text encoders as discriminators rather than generators,'' in \emph{International Conference on Learning Representations}, 2020. [Online]. Available: \url{https://openreview.net/forum?id=r1xMH1BtvB}
\BIBentrySTDinterwordspacing

\bibitem{liang2023holistic}
\BIBentryALTinterwordspacing
P.~Liang, R.~Bommasani, T.~Lee, D.~Tsipras, D.~Soylu, M.~Yasunaga, Y.~Zhang, D.~Narayanan, Y.~Wu, A.~Kumar, B.~Newman, B.~Yuan, B.~Yan, C.~Zhang, C.~A. Cosgrove, C.~D. Manning, C.~Re, D.~Acosta-Navas, D.~A. Hudson, E.~Zelikman, E.~Durmus, F.~Ladhak, F.~Rong, H.~Ren, H.~Yao, J.~WANG, K.~Santhanam, L.~Orr, L.~Zheng, M.~Yuksekgonul, M.~Suzgun, N.~Kim, N.~Guha, N.~S. Chatterji, O.~Khattab, P.~Henderson, Q.~Huang, R.~A. Chi, S.~M. Xie, S.~Santurkar, S.~Ganguli, T.~Hashimoto, T.~Icard, T.~Zhang, V.~Chaudhary, W.~Wang, X.~Li, Y.~Mai, Y.~Zhang, and Y.~Koreeda, ``Holistic evaluation of language models,'' \emph{Transactions on Machine Learning Research}, 2023, featured Certification, Expert Certification. [Online]. Available: \url{https://openreview.net/forum?id=iO4LZibEqW}
\BIBentrySTDinterwordspacing

\bibitem{llmperf}
ray project, ``llmperf,'' 2023, \url{https://github.com/ray-project/llmperf}.

\end{thebibliography}
\bibliographystyle{IEEEtran}

\end{document}